\documentclass[11pt]{article}

\usepackage {graphicx}
\usepackage{amssymb}
\usepackage{amsmath}
\usepackage{amsbsy}
\usepackage{bm}
\usepackage{wasysym}
\usepackage{textcomp}
\usepackage {graphicx}
\usepackage[export]{adjustbox}
\usepackage[]{algorithm2e}
\usepackage[titletoc,title]{appendix}
\usepackage{setspace}

\usepackage[usenames]{xcolor}

\newtheorem{theorem}{Theorem}
\newtheorem{corollary}{Corollary}
\newtheorem{proposition}{Proposition}
\newtheorem{lemma}{Lemma}
\newtheorem{example}{Example}

\newtheorem{remark}{Remark}

\newtheorem{proc}{Procedure}

\newtheorem{assumption}{Condition}

\newcommand{\qed}{\quad\hbox{\vrule width 4pt height 6pt depth 1.5pt}}

\newcommand{\hlambda}{\widehat{\lambda}}
\newcommand{\hN}{\widehat{N}}

\newcommand{\hR}{\widehat{R}}
\newcommand{\hr}{\widehat{r}}
\newcommand{\hQ}{\widehat{Q}}
\newcommand{\convp}{\overset{p}{\to}}
\newcommand{\convd}{\overset{d}{\to}}

\newcommand{\itY}{\mathcal{Y}}

\newcommand{\itI}{\mathcal{I}}
\newcommand{\sumim}{\sum_{i=1}^m}

\begin{document}

\title{Weighted False Discovery Rate Control in Large-Scale Multiple Testing}
\author{Pallavi Basu$^1$, \, T. Tony Cai$^2$, \,  Kiranmoy Das$^3$, \, and \, Wenguang Sun$^4$}

\footnotetext[1]{Department of Statistics and Operations Research, Tel Aviv University. Research supported in part by funding from the European Research Council under the European Community's Seventh Framework Programme (FP7/2007-2013) / ERC grant agreement n\textsuperscript{o}[294519]-PSARPS.} 
\footnotetext[2]{Department of Statistics, The Wharton School, University of Pennsylvania. The research of Tony Cai was supported in part by NSF Grant DMS-1403708 and NIH Grant R01 CA127334.}
\footnotetext[3]{Interdisciplinary Statistical Research Unit, Indian Statistical Institute, Kolkata. }
\footnotetext[4]{Department of Data Sciences and Operations, University of Southern California. The research of Wenguang Sun was supported in part by NSF grant DMS-CAREER 1255406.} 

\maketitle

\begin{abstract}
The use of weights provides an effective strategy to incorporate prior domain knowledge in large-scale inference. This paper studies weighted multiple testing in a decision-theoretic framework. We develop oracle and data-driven procedures that aim to maximize the expected number of true positives subject to a constraint on the weighted false discovery rate. The asymptotic validity and optimality of the proposed methods are established. The results demonstrate that incorporating informative domain knowledge enhances the interpretability of results and precision of inference. Simulation studies show that the proposed method controls the error rate at the nominal level, and the gain in power over existing methods is substantial in many settings. An application to a genome-wide association study is discussed.
\end{abstract}

 \thispagestyle{empty}

\newpage

\section{Introduction}

In large-scale studies, relevant domain knowledge, such as external covariates, scientific insights and prior data, is often available alongside the primary data set. Exploiting such information in an efficient manner promises to enhance both the interpretability of scientific results and precision of statistical inference. In multiple testing, the hypotheses being investigated often become ``unequal'' in light of external information, which may be reflected by differential attitudes towards the relative importance of testing units or the severity of decision errors. The use of weights provides an effective strategy to incorporate informative domain knowledge in large-scale testing problems. 

In the literature, various weighting methods have been advocated for a range of multiple comparison problems. A popular scheme, referred to as the \emph{decision weights} or loss weights approach, involves modifying the error criteria or power functions in the decision process (Benjamini and Hochberg, 1997). The idea is to employ two sets of positive constants $\pmb a=\{a_i: i=1, \cdots, m\}$ and $\pmb b=\{b_i: i=1, \cdots, m\}$ to take into account the costs and gains of multiple decisions. Typically, the choice of the weights $\pmb a$ and $\pmb b$ reflects the degree of confidence one has toward prior beliefs and external information. It may also be pertinent to the degree of preference that one has toward the consequence of one class of erroneous/correct decisions over another class based on various economical and ethical considerations. For example, in the  spatial cluster analysis considered by Benjamini and Heller (2007), the weighted false discovery rate was used to reflect that a false positive cluster with larger size would account for a larger error. Another example arises from genome-wide association studies (GWAS), where prior data or genomic knowledge, such as prioritized subsets (Lin and Lee, 2012), allele frequencies (Lin et al., 2014)  and expression quantitative trait loci information (Li et al., 2013), can often help to assess the scientific plausibility of significant associations. To incorporate such information in the analysis, a useful strategy is to up-weight the gains for the discoveries in preselected genomic regions by modifying the power functions in respective testing units (P\~ena et al., 2011; Sun et al., 2015; He et al., 2015). We assume in this paper that the weights have been pre-specified by the investigator. This is a reasonable assumption in many practical settings. For example, weights may be assigned according to economical considerations (Westfall and Young, 1993), external covariates (Benjamini and Heller, 2007; Sun et al., 2015) and biological insights from prior studies (Xing et al., 2010).

We mention two alternative formulations for weighted multiple testing. One popular method, referred to as the \emph{procedural weights} approach, involves the adjustment of the $p$-values from individual tests. In GWAS, Roeder et al.~(2006) and Roeder and Wasserman (2009) proposed to utilize linkage signals to up-weight the $p$-values in preselected regions and down-weight the $p$-values in other regions. It was shown that the power to detect association can be greatly enhanced if the linkage signals are informative, yet the loss in power is small when the linkage signals are uninformative. Another useful weighting scheme, referred to as the \emph{class weights} approach, involves allocating varied test levels to different classes of hypotheses. For example, in analysis of the growth curve data (Box, 1950), Westfall and Young (1993, page 186) proposed to allocate a higher family-wise error rate (FWER) to the class of hypotheses related to the primary variable ``gain'' and a lower FWER to the secondary variable ``shape''.

We focus on the decision weights approach in the present paper. This weighting scheme is not only practically useful for a wide range of applications, but also provides a powerful framework that enables a unified investigation of various weighting methods. Specifically, the proposal in Benjamini and Hochberg (1997)  involves the modification of both the error rate and power function. The formulation is closely connected to the classical ideas in compound decision theory that aim to optimize the tradeoffs between the gains and losses when many simultaneous decisions are combined as a whole. Our theory reveals that if the goal is to maximize the power subject to a given error rate, then the modifications via decision weights would lead to improved multiple testing methods with sensible procedural weights or class weights, or both. For example, in GWAS, the investigators can up-weight the power functions for discoveries in genomic regions that are considered to be more scientific plausible or biologically meaningful; this would naturally up-weight the $p$-values in these regions and thus yield weighting strategies similar to those suggested by Roeder and Wasserman (2009). In large clinical trials, modifying the power functions for respective rejections at the primary and secondary end points would correspond to the allocation of varied test levels across different classes of hypotheses, leading to weighting strategies previously suggested by Westfall and Young (1993).

The false discovery rate (FDR; Benjamini and Hochberg, 1995) has been widely used in large-scale multiple testing as a practical and powerful error criterion. Following Benjamini and Hochberg (1997), we generalize the FDR to weighted false discovery rate (wFDR), and develop optimal procedures for wFDR control under the decision weights framework. We first construct an oracle procedure that maximizes the weighted power function subject to a constraint on the wFDR, and then develop a data-driven procedure to mimic the oracle and establish its asymptotic optimality. The numerical results show that the proposed method controls the wFDR at the nominal level, and the gain in power over existing methods is substantial in many settings. {Our optimality result marks a clear departure from existing works in the literature on covariate-assisted inference, which aims to incorporate the external information by deriving optimal procedural weights for $p$-values (Roeder and Wasserman, 2009; Roquain and van de Wiel, 2009) or constructing covariate-adjusted test statistics (Ferkingstad et al.~2008; Zablocki et al.~2014; Cai et al.~2016) subject to the unweighted FDR and power function.}

Our research also makes a novel contribution to the theory of optimal ranking in multiple testing. Conventionally, a multiple testing procedure operates in two steps: ranking the hypotheses according to their significance levels and then choosing a cutoff along the rankings. It is commonly believed that the rankings remain the same universally at all FDR levels. For example, the ranking based on $p$-values or adjusted $p$-values in common practice is invariant to the choice of the FDR threshold. The implication of our theory is interesting, for it claims that there does not exist a ranking that is universally optimal at all test levels. Instead, the optimal ranking of hypotheses depends on the pre-specified wFDR level. That is, the hypotheses may be ordered differently when different wFDR levels are chosen. This point is elaborated in Section A of the Supplementary Material. See also Roeder and Wasserman (2009) and Roquain and van de Wiel (2009), where the FDR level $\alpha$ is included in constructing the procedural weights.

The rest of the article is organized as follows. Section 2 discusses a general framework for weighted multiple testing. Sections 3 and 4 develop oracle and data-driven
wFDR procedures and establish their optimality properties. Practice guidelines and numerical results are presented in Section 5. Section 6 concludes the article with a discussion of related and future works. Proof of Theorem 1 is given in Section 7. Proofs of other theoretical results and additional discussions are provided in the Supplementary Material. 

\section{Problem Formulation}

This section discusses a decision weights framework for weighted multiple testing. We first introduce the model and notation and then discuss modified error criteria and power functions.

\subsection{Model and notation}

Suppose that $m$ hypotheses $H_1, \cdots, H_m$ are tested simultaneously based on observations $X_1, \cdots, X_m$. Let $\pmb\theta=(\theta_1, \cdots, \theta_m)\in\{0, 1\}^m$ denote the true state of nature, where $0$ and $1$ stand for null and non-null cases, respectively. Assume that observations $X_i$ are independent and distributed according to the following model  
\begin{equation}\label{rmix}
X_i|\theta_i \sim (1-\theta_i) F_{0i} + \theta_i F_{1i},
\end{equation}
where $F_{0i}$ and $F_{1i}$ are the null and non-null distributions for $X_i$, respectively. Denote by $f_{0i}$ and $f_{1i}$ the corresponding density functions. Suppose that the unknown states $\theta_i$ are Bernoulli ($p_i$) variables, where $p_i=P(\theta_i=1)$. The mixture density is denoted by $f_{\cdot i} = (1-p_i)f_{0i} + p_if_{1i}$. 

Consider the widely used random mixture model (Efron et al., 2001; Storey, 2002; Genovese and Wasserman, 2002)
\begin{equation}\label{rmix-homo}
X_i \sim F=(1-p) F_{0} + p F_{1}.
\end{equation}
This model, which assumes that all observations are identically distributed according to a common distribution $F$, can sometimes be unrealistic in applications. In light of domain knowledge, the observations are likely to have different distributions. For example, in the context of a brain imaging study, Efron (2008) showed that  the proportions of activated voxels are different for the front and back halves of a brain. In GWAS, certain genomic regions contain higher proportions of significant signals than other regions. In the adequate yearly progress study of California high schools (Rogosa, 2003), the densities of $z$-scores vary significantly from small to large schools. We shall develop theory and methodology for model (\ref{rmix}), which allows possibly different non-null proportions and densities and is applicable to a wide range of settings. 

The multi-group model considered in Efron (2008) and Cai and Sun (2009), which has been widely used in applications, is an important case of the general model (\ref{rmix}). The multi-group model assumes that the observations can be divided into $K$ groups. Let $\mathcal G_k$ denote the index set of the observations in group $k$, $k=1, \cdots, K$. For each $i\in \mathcal G_k$, $\theta_{i}$ is distributed as Bernoulli$(p_k)$, and $X_{i}$ follows a mixture distribution: 
\begin{equation}\label{multi-group}
(X_i|i \in \mathcal G_k) \sim f_{\cdot k} = (1-p_k)f_{0k} +p_kf_{1k},
\end{equation}
where $f_{0k}$ and $f_{1k}$ are the null and non-null densities for observations in group $k$. This model will be revisited in later sections. See also Ferkingstad et al.~(2008), Hu et al.~(2010) and Liu et al.~(2016) for related works on multiple testing with groups.

\subsection{Weighted error criterion and power function}

This section discusses a generalization of the FDR criterion in the context of weighted multiple testing. Denote the decisions for the $m$ tests by $\pmb\delta=(\delta_1, \cdots, \delta_m)\in\{0, 1\}^m$, where $\delta_i=1$ indicates that $H_i$ is rejected and $\delta_i=0$ otherwise. Let $a_i$ be the weight indicating the severity of a false positive decision. For example, $a_i$ is taken as the cluster size in the spatial cluster analyses conducted in Benjamini and Heller (2007) and Sun et al.~(2015). As a result, rejecting a larger cluster erroneously corresponds to a more severe decision error. {To incorporate domain knowledge in multiple testing, Benjamini and Hochberg (1997) defined the weighted FDR $\mbox{wFDR}_{\rm{BH}} = E\{Q(\bf a)\}$, where
\begin{equation}\label{wFDR:BH}
  Q(\bf a) = \begin{cases} 
   \frac{\sum\limits_{i=1}^m a_i (1-\theta_i)\delta_i}{\sum\limits_{i=1}^m a_i\delta_i}, & \text{if } \sum\limits_{i=1}^m a_i\delta_i > 0, \\
   0,       & \text{otherwise.}
  \end{cases}
\end{equation}
We consider a slightly different version of the wFDR:
\begin{equation}\label{wFDR}
\mbox{wFDR} =  \frac{E\left\{\sum\limits_{i=1}^m a_i (1-\theta_i)\delta_i\right\}}{E\left(\sum\limits_{i=1}^m a_i\delta_i\right)}.
\end{equation}
In Section D of the Supplementary Material, we show that the definitions in \eqref{wFDR:BH} and \eqref{wFDR} are asymptotically equivalent. The main consideration of using \eqref{wFDR} is to facilitate our theoretical derivations and obtain exact optimality results. 

To assess the effectiveness of different wFDR procedures, we define the expected number of true positives
\begin{equation}
 \mbox{ETP} = E\left(\sum_{i=1}^m b_i \theta_i\delta_i\right),
\end{equation}
where $b_i$ is the weight indicating the power gain when $H_i$ is rejected correctly. The use of $b_i$ provides an effective scheme to incorporate domain knowledge. In GWAS, larger $b_i$ can be assigned to pre-selected genomic regions to reflect that the discoveries in these regions are more biologically meaningful. In spatial data analysis, correctly identifying a larger cluster that contains signal
may correspond to a larger $b_i$, indicating a greater gain. 

By combining the concerns on both the error criterion and power function, the goal in weighted multiple testing is to
\begin{equation}\label{WFDR-problem}
\mbox{\bf maximize the ETP subject to the constraint $\mbox{wFDR}\leq\alpha$}. 
\end{equation}
The optimal solution to (\ref{WFDR-problem}) is studied in the next section.

\section{Oracle Procedure for wFDR Control}

\setcounter{equation}{0}

The basic framework of our theoretical and methodological developments is outlined as follows. In Section 3.1, we assume that $p_i$, $f_{0i}$, and $f_{\cdot i}$ in the mixture model (\ref{rmix}) are known by an oracle and derive an oracle procedure that maximizes the ETP subject to a constraint on the wFDR. Connections to the literature is given in Section 3.2. In Section 4, we develop a data-driven procedure to mimic the oracle and establish its asymptotic validity and optimality.

\subsection{Oracle procedure}

The derivation of the oracle procedure involves two key steps: the first is to derive the optimal ranking of hypotheses and the second is to determine the optimal threshold along the ranking that exhausts the pre-specified wFDR level. We discuss the two issues in turn.

Consider model (\ref{rmix}). Define the local false discovery rate (Lfdr, Efron et al.~2001) as
\begin{equation}\label{Lfdr}
\mbox{Lfdr}_i= \frac{(1-p_i)f_{0i}(x_i)}{f_{\cdot i}(x_i)}. 
\end{equation}
The wFDR problem (\ref{WFDR-problem}) is equivalent to the following constrained optimization problem
\begin{equation}\label{WFDR-problem2}
\mbox{maximize $E \left\{ \sum\limits_{i=1}^m b_i \delta_i (1 - \mbox{Lfdr}_i) \right\}$ subject to $E \left\{ \sum\limits_{i=1}^m a_i \delta_i(\mbox{Lfdr}_i -\alpha) \right\}\leq 0$.} 
\end{equation}
Let $S^-=\{i: \mbox{Lfdr}_i\leq \alpha\}$ and $S^+=\{i: \mbox{Lfdr}_i>\alpha\}$. Then the constraint in (\ref{WFDR-problem2}) can be equivalently expressed as
\begin{equation}\label{WFDR-capacity}
E \left\{ \sum_{S^+} a_i \delta_i(\mbox{Lfdr}_i -\alpha) \right\} \leq E \left\{ \sum_{S^-} a_i \delta_i(\alpha-\mbox{Lfdr}_i) \right\}.
\end{equation}
Consider an optimization problem which involves packing a knapsack with a capacity given by the right hand side of equation (\ref{WFDR-capacity}). Every available object has a known value and a known cost (of space). Clearly rejecting a hypothesis in $S^-$ is always beneficial as it allows the capacity to expand, and thus promotes more discoveries. Hence the key issue boils down to how to efficiently utilize the capacity (after all hypotheses in $S^-$ are rejected) to make as many discoveries as possible in $S^+$. Observe that each rejection in $S^+$ would simultaneously increase the power and decrease the capacity. Intuitively, we should sort all hypotheses in $S^+$ in an decreasing order of the value to cost ratio (VCR). Equations (\ref{WFDR-problem2}) and (\ref{WFDR-capacity}) suggest that 
\begin{equation}\label{VCR}
\mbox{VCR}_i=\frac{b_i(1-\mbox{Lfdr}_i)}{a_i(\mbox{Lfdr}_i - \alpha)}. 
\end{equation}
To maximize the power, the ordered hypotheses are rejected sequentially until maximum capacity is reached.

The above considerations motivate us to consider the following class of decision rules $\pmb\delta^*(t)=\{\delta_i^*(t): i=1, \cdots, m\}$, where   
\begin{equation}\label{Ranking}
\delta_i^* (t) =
  \begin{cases}
  1, & \text{   if   } b_i(1-\mbox{Lfdr}_i) > t a_i(\mbox{Lfdr}_i - \alpha), \\
  0, & \text{   if   } b_i(1-\mbox{Lfdr}_i) \leq t a_i(\mbox{Lfdr}_i - \alpha).
   \end{cases}
\end{equation}
We briefly explain some important operational characteristics of testing rule (\ref{Ranking}). First, if we let $t>0$, then the equation implies that $\delta_i^*(t)=1$ for all $i\in S^-$; hence all hypotheses in $S^-$ are rejected as desired. (This explains why the VCR is not used directly in (\ref{Ranking}), given that the VCR is not meaningful in $S^-$.) Second, a solution path can be generated as we vary $t$ continuously from large to small. Along the path $\pmb\delta^*(t)$ sequentially rejects the hypotheses in $S^+$ according to their ordered VCRs. Denote by $H_{(1)}, \cdots, H_{(m)}$ the hypotheses sequentially rejected by $\pmb\delta^*$. (The actual ordering of the hypotheses within $S^-$ does not matter in the decision process since all are always rejected.)  

The next task is to choose a cutoff along the ranking to achieve exact wFDR control. The difficulty is that the maximum capacity may not be attained by a sequential rejection procedure. To exhaust the wFDR level, we  permit a randomized decision rule. Denote the Lfdr values and the weights corresponding to $H_{(i)}$ by $\mbox{Lfdr}_{(i)}$, $a_{(i)}$, and $b_{(i)}$. Let 
\begin{equation}\label{capacity}
C(j) = \sum_{i=1}^j a_{(i)} (\mbox{Lfdr}_{(i)}- \alpha)
\end{equation}
denote the capacity up to $j$th rejection. According to the constraint in equation \eqref{WFDR-problem2}, we choose $k=\max\{j: C(j)\leq 0\}$ so that the capacity is not yet reached when $H_{(k)}$ is rejected but would just be exceeded if $H_{(k+1)}$ is rejected. The idea is to split the decision point at $H_{(k+1)}$ by randomization. 

Let $U$ be a Uniform $(0, 1)$ variable that is independent of the truth, the observations, and the weights. Define 
\[
t^*=\frac{b_{(k+1)}\left(1-\mbox{Lfdr}_{(k+1)}\right)}{a_{(k+1)}\left(\mbox{Lfdr}_{(k+1)}- \alpha\right)} 
\quad \mbox{ and } \quad 
p^* = - \frac{C(k)}{C(k + 1)-C(k)}.
\]
Let $\mathcal I_A$ be an indicator, which takes value 1 if event $A$ occurs and 0 otherwise. We propose the \emph{oracle decision rule} $\pmb\delta_{OR}=\{\delta_{OR}^i: i=1, \cdots, m\}$, where
\begin{equation} 
\label{eq:oracle proc}
\delta_{OR}^i =
  \begin{cases}
  1 & \text{   if   } b_i(1-\mbox{Lfdr}_i) > t^* a_i(\mbox{Lfdr}_i - \alpha), \\ 
  0 & \text{   if   } b_i(1-\mbox{Lfdr}_i) < t^* a_i(\mbox{Lfdr}_i - \alpha), \\
  \mathcal I_{U < p^*} & \text{ if } b_i(1-\mbox{Lfdr}_i) = t^* a_i(\mbox{Lfdr}_i - \alpha).
   \end{cases}
\end{equation}

\begin{remark}{\rm
The randomization step is only employed for theoretical considerations to enforce the wFDR to be exactly $\alpha$. Thus the optimal power can be effectively characterized. Moreover, only a single decision point at $H_{(k+1)}$ is randomized, which has a negligible effect in large-scale testing problems. We do not pursue randomized rules for the data-driven procedures developed in later sections.  
}
\end{remark}

Let $\mbox{wFDR}(\pmb\delta)$ and $\mbox{ETP}(\pmb\delta)$ denote the wFDR and ETP of a decision rule $\pmb\delta$, respectively. Theorem \ref{thm:oracle_optimality} shows that the oracle procedure (\ref{eq:oracle proc}) is valid and optimal for wFDR control.

\begin{theorem}\label{thm:oracle_optimality}
 Consider  model (\ref{rmix}) and oracle procedure $\pmb\delta_{OR}$ defined in (\ref{eq:oracle proc}). Let $\mathcal D_\alpha$ be the collection of decision rules such that for any $\pmb\delta\in \mathcal D_\alpha$, $\mbox{wFDR}(\pmb\delta)\leq\alpha$. Then we have 
 
 \vspace{-15pt}
 \begin{itemize}
 \item[{\rm (i).}] $\mbox{wFDR}(\pmb \delta_{OR})=\alpha$. 
 \item[{\rm (ii).}] $\mbox{ETP}(\pmb\delta_{OR})\geq \mbox{ETP}(\pmb\delta)$ for all $\pmb\delta\in\mathcal D_\alpha$. 
 \end{itemize}
\end{theorem}

\subsection{Comparison with the optimality results in Spj$\o$tvoll (1972),  Benjamini and Hochberg (1997) and Storey (2007)} \label{pfer_control_section}

Spj$\o$tvoll (1972) showed that the likelihood ratio (LR) statistic 
\begin{equation}\label{T_LR}
T_{LR}^i=\frac{f_{0i}(x_i)}{f_{1i}(x_i)}
\end{equation}
is optimal for the following multiple testing problem 
\begin{equation}\label{IHTER-unweighted}
\mbox{ maximize $E_{\cap H_{1i}}  \left (\sum\limits_{i = 1}^m \delta_i  \right)$  subject to $E_{\cap H_{0i}} \left \{\sum\limits_{i = 1}^m \delta_i \right\} \leq \alpha$,}
\end{equation}
where $\cap H_{0i}$ and $\cap H_{1i}$ denote the intersections of the nulls and non-nulls, respectively. The error criterion $E_{\cap H_{0i}} \left \{\sum_i a_i \delta_i \right\}$ is referred to as the intersection tests error rate (ITER). 
A weighted version of problem (\ref{IHTER-unweighted}) was considered by Benjamini and Hochberg (1997), where the goal is  to
\begin{equation}\label{IHTER}
\mbox{ maximize $E_{\cap H_{1i}}  \left (\sum\limits_{i = 1}^m b_i \delta_i  \right)$  subject to $E_{\cap H_{0i}} \left \{\sum\limits_{i = 1}^m a_i \delta_i \right\} \leq \alpha$.}
\end{equation}
The optimal solution to (\ref{IHTER}) is given by the next proposition.

\begin{proposition} (Benjamini and Hochberg, 1997). Define the weighted likelihood ratio (WLR) 
\begin{equation}\label{T_IT}
T_{IT}^i=\frac{a_if_{0i}(x_i)}{b_if_{1i}(x_i)}. 
\end{equation}
Then the optimal solution to (\ref{IHTER}) is a thresholding rule of the form $\delta_{IT}^i=(T_{IT}^i<t_\alpha)$, where $t_\alpha$ is the largest threshold that controls the weighted ITER at level $\alpha$.
\end{proposition}
 
The ITER is very restrictive in the sense that the expectation is taken under the conjunction of the null hypotheses. The ITER is inappropriate for mixture model (\ref{rmix}) where a mixture of null and non-null hypotheses are tested simultaneously. To extend intersection tests to multiple tests, define the per family error rate (PFER) as
\begin{equation}\label{PFER-MHT}
\mbox{PFER}(\pmb\delta)= E\left\{\sum_{i=1}^m a_i (1-\theta_i)\delta_i\right\}.
\end{equation} 
The power function should be modified correspondingly. Therefore the goal is to  
\begin{equation}\label{PFER-problem}
\mbox{ maximize $E \left (\sum\limits_{i = 1}^m b_i \theta_i\delta_i  \right)$  subject to $E \left \{\sum\limits_{i = 1}^m a_i (1-\theta_i)\delta_i \right\} \leq \alpha$.}
\end{equation}
The key difference between the ITER and PFER is that the expectation in (\ref{PFER-MHT}) is now taken over all possible combinations of the null and non-null hypotheses. The optimal PFER procedure is given by the next proposition.  

\begin{proposition} \label{prop:pfer} Consider model (\ref{rmix}) and assume continuity of the LR statistic. Let $\mathcal D_{PF}^\alpha$ be the collection of decision rules such that for every $\pmb\delta\in\mathcal D_{PF}^\alpha$, $\mbox{PFER}(\pmb\delta)\leq\alpha$. Define the weighted posterior odds (WPO)
\begin{equation}\label{T_PF}
T_{PF}^i =\frac{a_i(1-p_i)f_{0i}(x_i)}{b_ip_if_{1i}(x_i)}.
\end{equation}
Denote by $Q_{PF}(t)$ the PFER of $\delta_{PF}^i=I(T_{PF}^i<t)$. Then the oracle PFER procedure is $\pmb\delta_{PF}=(\delta_{PF}^i: i=1, \cdots, m)$, where
$  
\delta_{PF}^i=I(T_{PF}^i<t_{PF}) \; \mbox{and} \; t_{PF}=\sup\{t: Q_{PF}(t)\leq \alpha\}.  
$
This oracle rule satisfies: 
\vspace{-15pt}
\begin{itemize}
\item[{\rm (i).}] $\mbox{ETP}(\pmb\delta_{PF})=\alpha$. 
\item[{\rm (ii).}] $\mbox{ETP}(\pmb\delta_{PF})\geq \mbox{ETP}(\pmb\delta)$ for all $\pmb\delta\in \mathcal D_{PF}^\alpha$. 
\end{itemize}
\end{proposition}

{Storey (2007) proposed the optimal discovery procedure (ODP) that aims to maximize the ETP subject to a constraint on the expected number of false positives (EFP). The ODP extends the optimality result in Spj$\o$tvoll (1972) from the intersection tests to multiple tests. However, the formulation of ODP does not incorporate weights. As a result, the ODP procedure is a symmetric rule where all hypotheses are exchangeable. Symmetric rules are in general not suitable for weighted multiple testing where it is desirable to incorporate   external information and treat the hypotheses differently.  
}




Our formulation (\ref{WFDR-problem}) modifies the conventional formulations in (\ref{IHTER}) and (\ref{PFER-problem}) to the multiple testing situation with an FDR type criterion. These modifications lead to methods that are more suitable for large-scale scientific studies. The oracle procedure (\ref{eq:oracle proc}) uses the VCR (\ref{VCR}) to rank the hypotheses and is an asymmetric rule. The VCR, which optimally combines the decision weights, significance measure (Lfdr) and test level $\alpha$, produces a more powerful ranking than the WPO (\ref{T_PF}) in the wFDR problem. Section A in the supplementary material provides a detailed discussion on the ranking issue.

\section{Data-Driven Procedures and Asymptotics}

\setcounter{equation}{0}
The oracle procedure (\ref{eq:oracle proc}) cannot be implemented in practice since it relies on unknown quantities such as $\mbox{Lfdr}_i$ and $t^*$. This section develops a data-driven procedure to mimic the oracle. We first propose a test statistic to rank the hypotheses and discuss  related estimation issues. A step-wise procedure is then derived to determine the best cutoff along the ranking. Finally, asymptotic results on the validity and optimality of the proposed procedure are presented.

\subsection{Proposed test statistic and its estimation}

The oracle procedure utilizes the ranking based on the VCR (\ref{VCR}). However, the VCR is only meaningful for the tests in $S^+$ and becomes problematic when both $S^-$ and $S^+$ are considered.  Moreover, the VCR could be unbounded, which would lead to difficulties in both numerical implementations and technical derivations. We propose to rank the hypotheses using the following statistic (in increasing values)
\begin{equation}\label{R-stat}
R_i = \frac{a_i(\mbox{Lfdr}_i - \alpha)}{b_i(1 - \mbox{Lfdr}_i) + a_i|\mbox{Lfdr}_i - \alpha|}.
\end{equation}
As shown in the next proposition,  $R_i$ always ranks hypotheses in $S^-$ higher than hypotheses in $S^+$ (as desired), and yields the same ranking as that by the VCR (\ref{VCR}) for hypotheses in $S^+$. The other drawbacks of VCR can also be overcome by $R_i$: $R_i$ is always bounded in the interval $[-1,1]$ and is a continuous function of the $\mbox{Lfdr}_i$. 
\begin{proposition}\label{prop:ranking} 
(i) The rankings generated by the decreasing values of VCR (\ref{VCR}) and increasing values of $R_i$ (\ref{R-stat}) are the same in both $S^-$ and $S^+$. (ii) The ranking based on increasing values of $R_i$ always puts hypotheses in $S^-$ ahead of hypotheses in $S^+$. 
\end{proposition}

Next we discuss how to estimate $R_i$; this involves the estimation of the Lfdr statistic (\ref{Lfdr}), which has been studied extensively in the multiple testing literature. We give a review of related methodologies. If all observations follow a common mixture distribution (\ref{rmix-homo}), then we can first estimate the non-null proportion $p$ and the null density $f_{0}$ using the methods in Jin and Cai (2007), and then estimate the mixture density $f$ using a standard kernel density estimator (e.g.~Silverman, 1986). If all observations follow a multi-group model (\ref{multi-group}), then we can apply the above estimation methods to separate groups to obtain corresponding estimates $\hat p_k$, $\hat f_{0k}$, and $\hat f_{\cdot k}$, $k=1, \cdots, K$. The theoretical properties of these estimators have been established in Sun and Cai (2007) and  Cai and Sun (2009). In practice, estimation problems may arise from more complicated models. Related theories and methodologies have been studied in Storey (2007), Ferkingstad et al.~(2008), and Efron (2008, 2010); theoretical supports for these estimators are yet to be developed. 

The estimated Lfdr value for $H_i$ is denoted by $\widehat{\mbox{Lfdr}}_i$. By convention, we take $\widehat{\mbox{Lfdr}}_i=1$ if $\widehat{\mbox{Lfdr}}_i>1$. This modification only facilitates the development of theory and has no practical effect on the testing results (since rejections are essentially only made for small $\widehat{\mbox{Lfdr}}_i$'s). The ranking statistic $R_i$ can therefore be estimated as
\begin{equation} \label{dd test stat}
\widehat{R}_i = \frac{a_i(\widehat{\mbox{Lfdr}}_i - \alpha)}{b_i(1 - \widehat{\mbox{Lfdr}}_i) + a_i|\widehat{\mbox{Lfdr}}_i - \alpha|}.
\end{equation}
The performance of the data driven procedure relies on the accuracy of the estimate $\widehat{\mbox{Lfdr}}_i$; some technical conditions are discussed in the next subsection. 
The finite sample performance of different Lfdr estimates are investigated in Section E.4 of the Supplementary Material.

\subsection{Proposed testing procedure and its asymptotic properties}

Consider $\widehat R_i$ defined in (\ref{dd test stat}). Denote by $\widehat{N}_i = a_i(\widehat{\mbox{Lfdr}}_i - \alpha)$ the estimate of excessive error rate when $H_i$ is rejected. Let $\widehat R_{(1)}, \cdots, \widehat R_{(m)}$ be the ordered test statistics (in increasing values). The hypothesis and estimated excessive error rate corresponding to $\widehat R_{(i)}$ are denoted by $H_{(i)}$ and $\widehat N_{(i)}$. The idea is to choose the largest cutoff along the ranking based on $\widehat R_i$ so that the maximum capacity is reached. Motivated by the constraint in (\ref{WFDR-problem2}), we propose the following step-wise procedure.

\begin{proc}\label{dd proc} (wFDR control with general weights). Rank hypotheses according to $\widehat R_i$ in increasing values.
Let $k=\max\left\{j: \sum\limits_{i=1}^j \widehat{N}_{(i)} \leq 0\right\}.$ Reject $H_{(i)}$, for $i = 1, \ldots, k$.
\end{proc}
It is important to note that in Procedure \ref{dd proc}, $\widehat R_i$ is used in the ranking step whereas $\widehat N_i$ (or a weighted transformation of $\widehat{\mbox{Lfdr}}_i$) is used in the thresholding step. The ranking by $\widehat{\mbox{Lfdr}}_i$ is in general different from that by $\widehat R_i$. In some applications where the weights are proportional, i.e. $\pmb a =c \cdot \pmb b$ for some constant $c>0$, then the rankings by $\widehat R_i$ and $\widehat{\mbox{Lfdr}}_i$ are identical. Specifically $\widehat R_i$ is then monotone in $\widehat{\mbox{Lfdr}}_i$. Further, choosing the cutoff based on $\widehat N_i$ is equivalent to that of choosing by a weighted $\widehat{\mbox{Lfdr}}_i$. This leads to an Lfdr based procedure (Sun et al., 2015), which can be viewed as a special case of Procedure \ref{dd proc}.
\begin{proc}\label{dd proc2} (wFDR control with proportional weights). Rank hypotheses according to $\widehat{\mbox{Lfdr}}_i$ in increasing values. Denote the hypotheses and weights corresponding to $\widehat{\mbox{Lfdr}}_{(i)}$ by $H_{(i)}$ and $a_{(i)}$. 
Let $$
k = \max \left\{ j: \left(\sum_{i = 1}^j a_{(i)} \right)^{-1} \sum_{i = 1}^j a_{(i)} \widehat{\mbox{Lfdr}}_{(i)}\leq \alpha \right\}.$$ Reject $H_{(i)}$, for $i = 1, \ldots, k$. 
\end{proc}



Next we investigate the asymptotic performance of Procedure \ref{dd proc}. We first give some regularity conditions for the weights. Our theoretical framework requires that the decision weights must be obtained from external sources such as prior data, biological insights, or economical considerations. In particular, \emph{the observed data $\{X_i: i=1, \cdots, m\}$ cannot be used to derive the weights}. The assumption is not only crucial in theoretical developments, but also desirable in practice (to avoid using data twice). Therefore given the domain knowledge, the decision weights do not depend on observed values. Moreover, a model with random (known) weights is employed for technical convenience, as done in Genovese et al.~(2006) and Roquain and van de Wiel (2009). We assume that the weights are independent with each other across testing units. Formally, denote $e_i$ the external domain knowledge for hypothesis $i$, we require the following condition.

\begin{assumption} \label{cons:wts1} 
(i) $(a_i, b_i|X_i, \theta_i, e_i) \overset{d}{\sim} (a_i, b_i|e_i)$ for $1\leq i\leq m$. (ii) $(a_i, b_i)$ and $(a_j, b_j)$  are independent for $i\neq j$. 
\end{assumption}
In weighted multiple testing problems, the analysis is always carried out in light of the external information $e_i$ implicitly. The notation of conditional distribution on $e_i$ will be suppressed when there is no ambiguity. In practice, the weights $a_i$ and $b_i$ are usually bounded. We need a weaker condition in our theoretical analysis.

\begin{assumption} \label{cond:wts2} (Regularity conditions on the weights.)
Let $C$ and $c$ be two positive constants. $E(\sup_i a_i) = o(m), E(\sup_i b_i)= o(m),  E(a_i^4) \leq C$,  and $\min\{E(a_i), E(b_i)\} \geq c$.
\end{assumption}

A consistent Lfdr estimate is needed to ensure the large-sample performance of the data-driven procedure. Formally, we need the following condition. 

\begin{assumption}\label{cond:estimates} $\widehat{\mbox{Lfdr}}_i - \mbox{Lfdr}_i = o_P(1)$. Also, $\widehat{\mbox{Lfdr}}_i \convd \mbox{Lfdr}$, where $\mbox{Lfdr}$ is an independent copy of $\mbox{Lfdr}_i$.
\end{assumption}

\begin{remark}{\rm
Condition 3 is a reasonable assumption in many applications. We give a few important scenarios where Condition 3 holds. Suppose we observe $z$-values from random mixture model
$$
Z_i\sim N(\mu_i, \sigma_i^2), \; (\mu_i, \sigma_i)\sim F(\mu, \sigma),
$$
where $(\mu_i, \sigma_i)=(\mu_0, \sigma_0)$ if $\theta_i=0$, $(\mu_i, \sigma_i)\neq (\mu_0, \sigma_0)$ if $\theta_i=1$, and $F(\mu, \sigma)$ is a general bivariate distribution. Let $p$ denote the proportion of non-null cases. Then $\hat{p}_{jc}$, the estimator proposed in Jin and Cai (2007), satisfies $\hat{p}_{jc}\xrightarrow{p} p$ under mild regularity conditions. 
Moreover, it is known that the kernel density estimator satisfies $E\|\hat{f}-f\|^2\rightarrow 0$. It follows from Sun and Cai (2007) that Condition 3 holds when the above estimators are used. If the null distribution is unknown, then we can use the method in Jin and Cai (2007) to estimate the null parameters $(\mu_0, \sigma_0)$. Then under certain regularity conditions, we can show that $E\|\hat{f}_{0}-f_{0}\|^2\rightarrow 0$, and Condition 3 holds.
For the multi-group model (\ref{multi-group}),  let $\hat p_k$, $\hat f_{k0}$, and $\hat f_k$ be estimates of $p_k$, $f_{k0}$, and $f_k$ such that $\hat{p}_k\xrightarrow{p} p_k$, $E\|\hat{f}_{k0}-f_{k0}\|^2\rightarrow 0$, $E\|\hat{f}_k-f_k\|^2\rightarrow 0$, $k=1, \cdots, K$. Let $\widehat{\mbox{Lfdr}}_i={(1-\hat{p}_k)\hat{f}_{0k}(x_{i})}/{\hat{f}_k(x_{i})}$ if $i\in\mathcal G_k$. It follows from Cai and Sun (2009) that Condition 3 holds when we apply Jin and Cai's estimators and standard kernel estimates to the groups separately.
}
\end{remark}

The oracle procedure (\ref{eq:oracle proc}) provides an optimal benchmark for all wFDR procedures. The next theorem establishes the asymptotic validity and optimality of the data-driven procedure by showing that the wFDR and ETP levels of the data-driven procedure converge to the oracle levels  as $m\rightarrow \infty$.  

\begin{theorem}\label{fdr control}  
Assume Conditions 1-3 hold. Denote by $\mbox{wFDR}_{DD}$ the wFDR level of the data-driven procedure (Procedure 1). Let $ETP_{OR}$ and $ETP_{DD}$ be the ETP levels of the oracle procedure \eqref{eq:oracle proc} and data-driven procedure, respectively. Then we have 

\vspace{-15pt}
\begin{itemize}
\item[{\rm (i).}] $\mbox{\rm wFDR}_{DD} = \alpha + o(1).$
\item[{\rm (ii).}]  $\mbox{\rm ETP}_{DD} / \mbox{\rm ETP}_{OR}=1 + o(1).$ 
\end{itemize}
\end{theorem}

\begin{corollary}
 Suppose we choose $a_i=1$ for all $i$. Then under the conditions of the theorem, our data-driven procedure controls the unweighted FDR at level $\alpha+o(1)$.  
\end{corollary}

\setcounter{equation}{0}

\section{Practical Issues and Numerical Results}

{

The wFDR framework provides a useful approach to integrate  domain knowledge in multiple testing. However, the weights must be chosen with caution to avoid improper manipulation of results. Under our formulation, we assume that the weights are {pre-specified}, while stressing that the task of assigning ``correct'' weights is a critical issue in practice. This section first states a few  practical guidelines in weighted FDR analysis, then describes general strategies in choosing weights for a range of application scenarios. Finally we illustrate the proposed methodology via an application to GWAS.

\subsection{Practical guidelines and examples on choosing weights}

We first state a few important practical guidelines to ensure that the weighted FDR analysis is conducted properly. The first guideline is in particular important for a valid FDR control.

\begin{enumerate}

 \item 
The weights must be ``external'' in the sense that they do not depend on the primary data such as the $z$-values or $p$-values on which the multiple tests are performed. As a practical guideline, we recommend ``choosing external weights before seeing the data'' as a standard rule in all weighted FDR analyses.  
 
 \item The choice of weights requires scientific motivations or economic considerations. For a statistical analysis involving any weighting schemes, the procedure for obtaining weights must be specified beforehand in the analysis plan, and needs be disclosed and justified carefully.

 \item To prevent researchers from setting weights to find desired significant results, extreme weights are in general not acceptable. It is recommended to use moderate weights to incorporate domain knowledge and carry out a sensitivity analysis afterwards to avoid manipulations of results. 
 
\end{enumerate}

Next we discuss a range of scenarios that naturally give rise to weighted FDR analyses. We use these examples to illustrate how to assign weights properly in light of various economic and scientific considerations. Moreover, we emphasize how the above practical guidelines can be implemented in respective contexts to avoid manipulations of results. 

\noindent\textbf{Example 1. Spatial cluster analysis}. In multiple testing, relevant domain knowledge or external data can be exploited to form more meaningful testing units such as sets or clusters. For example, in a spatial setting (Pacifico et al.~2004; Benjamini and Heller 2007; Sun et al.~2015), individual locations can be aggregated into clusters to increase the signal to noise ratio and scientific interpretability. It is natural to consider weighted error rates when clusters are heterogeneous. We state a few guidelines for the choice of weights.  
\begin{description}
 \item (i) The weights should be pre-specified by experts to reflect various practical and economic considerations based on relevant spatial covariates such as cluster sizes, population densities, poverty rates, and urban vs. rural regions. 
 
 \item (ii) In cluster-wise analyses, weights may be used to modify both the error rate and power function. For example, symmetric weights (i.e.~$a_i=b_i$ for all $i$) may be assigned to reflect that the gains and losses in the decision process are proportional. A careful investigation of the optimal ranking of clusters (Section A in the Supplementary Material) is helpful to improve the power of analysis. 
 
\end{description}
 
\noindent\textbf{Example 2. Multiple endpoints clinical trials.} In comparing treatments with multiple endpoints (e.g.~Dmitrienko et al.~2003), the error rates may be modified by assigning different weights to primary and secondary end points; this promises to increase the power in discovering more important clinical findings (Young and Gries 1984; Westfall and Young 1993).
Benjamini and Cohen (2016) discussed two weighting regimes in clinical trials: the first regime requires that any primary endpoint carries a larger weight than any secondary one; and the second regime in addition requires that the combined weight of the primary endpoints is greater than the combined weight of the secondary ones. To avoid manipulations of results, Benjamini and Cohen (2016) suggested that a weighted FDR analysis should obey the following rules: 
\begin{description}
 \item (i) The tradeoffs between primary and secondary end points should be evaluated prior to conducting the trial.
 \item (ii) The weights should be given before seeing the data. 
 \item (iii) The details of obtaining weights should be revealed in the trial protocol. 
\end{description}
 
\noindent\textbf{Example 3. Prioritized subset analysis.} In GWAS data analysis, finding disease-causing SNPs in the immense data is very challenging. Weighted FDR analysis provides a powerful approach to integrate genomic knowledge in inference with prioritized SNP subsets. For example, Roeder et al.~(2006) and Zablocki et al.~(2014) assigned higher weights to pre-selected genomic regions to increase the power in detecting disease-associated SNPs that are more biologically plausible. In large-scale multiple testing with prioritized subset, we have the following requirements regarding the choice of weights:

\begin{description}

 \item (i). It is not allowed to use the primary data (e.g.~significant levels of disease association) to derive weights; such practice would distort the null distribution of $p$-values and lead to the failure of the wFDR procedure.  

 \item (ii). The weights must be obtained based on either prior knowledge regarding the biological importance of an association, or external knowledge such as the linkage disequilibrium (LD) correlations between SNPs. This requirement can effectively avoid selection bias because the weights are conditionally independent from the significance levels of SNP associations (cf.~our Condition 1). See also Section 5.(a) in Benjamini et al.~(2009) for related discussions. 
 
 \item (iii). To avoid further aggravating the multiplicity issue in large-scale testing problems, the weights should only be used to modify the power function. The unweighted FDR criterion is recommended to remain unchanged. This ensures greater power to reject more biologically relevant hypotheses without increasing the FDR (cf.~our Corollary 1). See also Theorem 1 in Genovese et al.~(2006) and Section 5.(a) in Benjamini et al.~(2009) for related theoretical results and discussions.

\end{description} 
}

\subsection{Decision weights vs. procedural weights}

The formulation \eqref{WFDR-problem} is different from the ``procedural weights'' approach, for example, in Roeder et al.~(2006) and Zablocki et al.~(2014), where the goal is to exploit external information by constructing weights to increase the total number of rejections (covariate-assisted inference). The decision weights framework integrates the external information in a very different way; the goal is to improve the interpretability of results instead of statistical power (although the power may be gained as a byproduct, see Section E.5 in the Supplementary Material). Under the wFDR framework, the hypotheses are of ``unequal'' importance in light of domain knowledge. For instance, in spatial cluster analysis, a false positive cluster with larger size would account for a larger error; hence it is not sensible to use the conventional FDR definition that treats all testing units equally. The wFDR framework is different from those in Genovese et al.~(2006), Ferkingstad et al.~(2008) and Cai et al.~(2016), where the gains and losses in various decisions are exchangeable, and conventional FDR and power definitions are suitable.

\subsection{Application to Framingham Heart Study (FHS)}

In this section, we implement the proposed method for analyzing a data set from Framingham Heart Study (Fox et al., 2007; Jaquish, 2007). A brief description of the study, the implementation of our methodology, and the results are discussed in turn.

The goal of the study is to decipher the genetic architecture behind the cardiovascular disorders for the Caucasians. Started in 1948 with nearly 5,000 healthy subjects, the study is currently in its third generation of the participants. The biomarkers responsible for the cardiovascular diseases, for e.g., body mass index (BMI), weight, blood pressure, and cholesterol level, were measured longitudinally. {Since the mutation SNPs (within the block of DNA) are commonly passed on to descendants, it is a standard practice to collect data over several generations. The specific goal of our study is to identify disease-associated SNPs in the second generation while paying more attention to those that remain stable over the first two generations. Identifying significant SNPs (or SNP blocks) that are preserved across generations can improve both biological plausibility and statistical replicability. In our study design, we use 399 subjects in the first generation to serve as the baseline to construct weights, and use the information from the 578 subjects in the second generation (with 310 males and 268 females) to conduct the primary analysis. It is important to note that the goal of our weighted FDR analysis is different from that in Roeder et al.~(2006) and Zablocki et al.~(2014), where prior likelihood of association with the phenotype is utilized for constructing ``procedural weights'' to improve the unweighted power. In contrast, we use decision weights to modify the power function so that discoveries in certain preselected regions receive a higher priority. The main consideration is not to increase the total number of discoveries, but to focus on findings are more relevant to our study goal (identifying stable mutations across generations). In our study it is not sensible to use the conventional power definition where the discoveries from all regions are treated equally.

We consider the BMI as the response variable and develop a dynamic model to detect the SNPs associated with the BMI. Let $Y_{i}(t_{ij})$ denote the response (BMI) from the $i$-th subject at time $t_{ij}$, $j=1,\ldots,T_i$. Consider the following  model for longitudinal traits:
\begin{equation} \label{dyn model}
 Y_{i}(t_{ij})=f(t_{ij})+\beta_k G_{ik}+\gamma_{i0}+\gamma_{i1}t_{ij}+\epsilon_i(t_{ij}),
  \end{equation}
 where $f(\cdot)$ is the general effect of time that is modeled by a polynomial function of suitable order, selected by AIC or BIC model selection criterion, $\beta_k$ is the effect of the $k$-th SNP on the response and $G_{ik}$ denotes the genotype of the $i$-th subject for the $k$-th SNP. We also consider the random intercepts and random slopes, denoted $\gamma_{0i}$ and $\gamma_{1i}$, respectively, for explaining the subject-specific response trajectories. A bivariate normal distribution for $\gamma_i=(\gamma_{0i}, \gamma_{1i})$ is assumed. Moreover, we assume that the residual errors are normally distributed with zero mean, and covariance matrix $\Sigma_i$ with an order-one auto-regressive structure. 
  We fit model (\ref{dyn model}) for each SNP and obtain the estimate of the genetic effect $\widehat{\beta}_k$. If we reject the null hypothesis $H_0:\beta_k=0$ vs.~$H_1:\beta_k \neq 0$, then we conclude that the $k$-th SNP has a significant association with the BMI. Since we have nearly 5 million SNPs, the false discovery rate needs to be controlled  for making scientifically meaningful inference. For each $k$, we take standardized $\widehat{\beta}_k$ as our $z$-scores and obtain the estimated ranking statistic $\widehat{R}_k$ as described in (\ref{dd test stat}). 

To incorporate external domain knowledge, we construct weights by utilizing the $p$-values obtained from first generation participants. This information is secondary in the sense that the first generation data are not directly used in multiple testing but are only employed to provide supplementary support in inference. In the FHS study, there are 399 subjects in the first generation for which we have obtained the trait values (such as BMI) and  genetic information on the same set of SNPs. Using the $p$-values obtained from the first generation data, we partition the SNPs into three groups: less than 0.001, between 0.001 and 0.01, and greater than 0.01. The groups are denoted $\mathcal G_j$, $j=1, 2, 3$. We implement the proposed wFDR method using the $p$-values computed from second generation participants, with group-wise weights $b_i=c_j$ if $i\in \mathcal G_j$. Note that $b_i$ are the weights in respective power functions with a larger $b_i$ indicating a higher priority or a stronger  belief. Since assigning a value to the biological importance of an association can be quite subjective, we propose to use different combinations of weights to investigate the sensitivity of testing results:
\begin{description}
  \item Setting 1: $a_i=1$ for all $i$, $b_i=c_j$ if $i\in G_j$, $(c_1, c_2, c_3)=(1, 1, 1)$;
  \item Setting 2: $a_i=1$ for all $i$, $b_i=c_j$ if $i\in G_j$, $(c_1, c_2, c_3)=(4, 2, 1)$;
  \item Setting 3: $a_i=1$ for all $i$, $b_i=c_j$ if $i\in G_j$, $(c_1, c_2, c_3)=(10, 5, 1)$.
\end{description}
Setting 1 corresponds to an unweighted FDR analysis. Settings 2 and 3 reflect our belief that the correct rejections from groups 1 and 2 are more ``powerful'' or more ``relevant'' when compared to those from group 3. This is sensible as one of the goals in the FHS study is to discover significant SNPs that remain \emph{stable} across generations. In our analysis, the weights $a_i=1$ are chosen for all SNPs; this ensures that \emph{the FDR will not be increased by weighting}. We only modify the weights $b_i$ in the power functions to reflect that discoveries in certain pre-selected regions are of greater importance. In practice $b_i$ may be derived from expert opinion to incorporate scientific or economic considerations. We stress that in GWAS, the details of the process in choosing weights, which requires a separate discussion or justification, should be revealed.

We implement the proposed wFDR method to select  most significant SNPs.  The results for the sensitivity analysis with different choices of weights are summarized in Table \ref{analysis-result.tab}. The table shows the groups sizes, the group-wise threshold levels, and the number of SNPs selected from different ``groups'' for each weight combination at two different FDR levels $\alpha = 0.10$ and $\alpha = 0.05$. 

\begin{table}\caption{Number of rejections and thresholds by varying choice of weights between groups.}\label{analysis-result.tab}
\begin{center}
\begin{tabular}{cc|cccccc}
\multicolumn{8}{c}{}\\
\hline
\multicolumn{8}{c}{Choice of Weights}\\
\hline
 Group &  Total & (1, 1, 1) & Threshold & (4, 2, 1) & Threshold & (10, 5, 1) & Threshold\\ 
\hline
\hline
\multicolumn{8}{c}{$\alpha = 0.10$}\\
\hline
\hline
$\mathcal G_1$ & 1043 & 3 & $6.69 \times 10^{-4}$ & 5 & $1.70 \times 10^{-3}$ & 53 & $3.17 \times 10^{-2}$\\
$\mathcal G_2$ & 5277 & 23 & $9.01 \times 10^{-4}$ & 33 & $1.52 \times 10^{-3}$ & 57 & $4.53 \times 10^{-3}$\\
$\mathcal G_3$ & 220392 & 2019 & $2.92 \times 10^{-3}$ & 1997 & $2.88 \times 10^{-3}$ & 1663 & $2.31 \times 10^{-3}$\\
Overall & 226712 & 2045 & -- & 2035 & -- & 1773 & --\\
\hline
\hline
\multicolumn{8}{c}{$\alpha = 0.05$}\\
\hline
\hline
$\mathcal G_1$ & 1043 & 1 & $5.22 \times 10^{-6}$ & 2 & $3.95 \times 10^{-4}$ & 4 & $7.43 \times 10^{-4}$\\
$\mathcal G_2$ & 5277 & 13 & $4.22 \times 10^{-4}$ & 19 & $5.20 \times 10^{-4}$ & 21 & $6.61 \times 10^{-4}$\\
$\mathcal G_3$ & 220392 & 159 & $1.17 \times 10^{-4}$ & 147 & $1.06 \times 10^{-4}$ & 131 & $1.01 \times 10^{-4}$\\
Overall & 226712 & 173 & -- & 168 & -- & 156 & --\\
\hline
\end{tabular}
\end{center}
\end{table}
 
From Table \ref{analysis-result.tab} we can see that the unweighted analysis [i.e.~the weights combination (1, 1, 1)] consistently selects higher number of SNPs than the other two weights combinations. Meanwhile, it selects fewer number of SNPs from groups 1 and 2. Here we have aimed to prioritize certain pre-selected regions and discover significant SNPs that are more biologically relevant. In our analysis, groups 1 and 2 are believed to be more ``informative.'' At $\alpha = 0.10$, it is interesting to note that the number of rejections from groups 1 and 2 are increased by several folds but the number of rejections from group 3 is only decreased by a small proportion.  

In Table 1, we also report the group-wise thresholds for the $p$-values. The value of the threshold is computed as the maximum of the $p$-values of the hypotheses rejected in the specific group and weight combination. 
We can see that the threshold for group 3 decreases in weights, which indicates that the rejection criteria become more stringent with higher weights. In contrast, the thresholds for groups 1 and 2 increase with weights, which allows for more discoveries in pre-selected regions. All these above observations are sensible and in agreement with the intuitions of our proposed methodology. }

\subsection{Simulation}

\setcounter{equation}{0}

In all simulation studies, we consider a two-point normal mixture model 
\begin{equation}\label{2p-nmix.model}
X_{i} \sim (1-p) N(0, 1) + pN(\mu,\sigma^2), i=1, \cdots, m.
\end{equation}
The nominal wFDR is fixed at $\alpha=0.10$. We consider the comparison of different methods under the scenario where there are two groups of hypotheses and within each group the weights are proportional. 

The proposed method (Procedure 1 in Section 4.2) is denoted by $\triangle$ DD. Other methods to be compared include: 
\begin{enumerate}
  \item The wFDR method proposed by Benjamini and Hochberg (1997); denoted by $\Box$ BH97. In simulations where $a_i=1$ for all $i$, BH97 reduces to the well-known step-up procedure in Benjamini and Hochberg (1995), denoted by BH95.
  \item A stepwise wFDR procedure, which rejects hypotheses along the WPO (\ref{T_PF}) ranking sequentially and stops at $k=\max\left\{j: \sum\limits_{i=1}^j \widehat{N}_{(i)} \leq 0\right\}$, with $\widehat{N}_{(i)}$ defined in Section 4.2. The method is denoted by $\circ$ WPO. Following similar arguments in the proof of Theorem 2, we can show that the WPO method controls the wFDR at the nominal level asymptotically. This is also verified by our simulation results. Meanwhile, we expect that the WPO method will be outperformed by the proposed method ($\triangle$ DD), which operates along the more efficient VCR ranking.
  \item The adaptive $z$-value method in Sun and Cai (2007), denoted by $+$ AZ. AZ is valid and optimal in the unweighted case but suboptimal in the weighted case. 
\end{enumerate}

To save space, this section only presents results on group-wise weights. Our setting is motivated by our application to GWAS, where the hypotheses can be divided into groups: those in preselected regions and those in other regions. It is desirable to assign varied weights to separate groups to reflect that the discoveries in preselected regions are more biologically meaningful. {In Section E.2 of the Supplementary Material, we compare our methods with existing methods using general weights $a_i$ and $b_i$ that are generated from probability distributions. We also provide additional numerical results such as the comparison of various wFDR definitions, the finite sample performance of Lfdr, and the impacts of weights on the power of different wFDR procedures.}

The first simulation study investigates the effect of weights. Consider two groups of hypotheses with group sizes $m_1 = 3000$ and $m_2 = 1500$.  In both groups, the non-null proportion is $p=0.2$. The null and non-null distributions are $N(0, 1)$ and $N(1.9, 1)$, respectively. We fix $a_i=1$ for all $i$. Hence BH97 reduces to the method proposed in Benjamini and Hochberg (1995), denoted by BH95. The wFDR reduces to the regular FDR, and all methods being considered are valid for FDR control. For hypotheses in group 1, we let $c_1 = a_{i}/b_{i}$. For hypotheses in group 2, we let $c_2=a_{i}/b_{i}$. We choose $c_1=3$ and vary $c_2$. Hence the weights are proportional within respective groups and vary across groups. The second simulation study investigates the impacts of signal strength. The results are presented in Section E.1 of the Supplementary material. 

In the simulation, we apply the four methods above to the simulated data set and obtain the wFDR and ETP levels by averaging the multiple testing results over 200 replications. In Figure \ref{figure1_plot}, we plot the wFDR levels and ETP of different methods as functions of $c_2$, which is varied over $[0.1,0.8]$. Panel (a) shows that all methods control the wFDR under the nominal level, and the BH97 method is conservative. Panel (b) shows that the proposed method dominates all existing methods. The proposed method is followed by the WPO method, which outperforms all unweighted methods (AZ and BH95) since $b_i$, the weights in the power function, are incorporated in the testing procedure. The BH97 (or BH95) has the smallest ETP. As $c_2$ approaches 1 or the weights $a_i$ and $b_i$ equalizes, the relative difference of the various methods (other than BH95) becomes less.

\begin{figure}
\centering
\includegraphics[max size={\textwidth}{\textheight}]{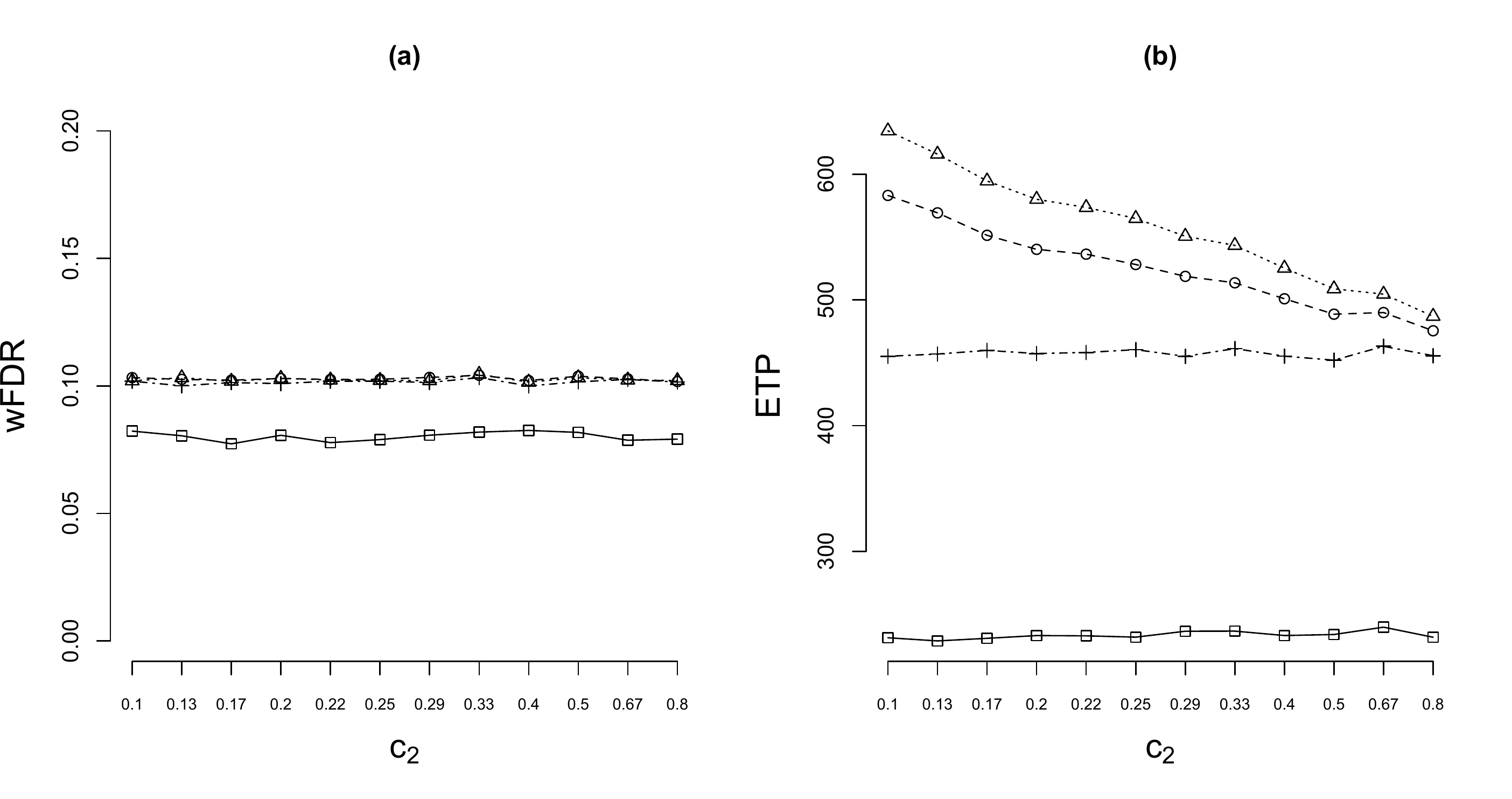} 
\caption{Comparison under group-wise weights: $\Box$ BH97 (or BH95), $\circ$ WPO, $\triangle$ DD (proposed), and $+$ AZ. The efficiency gain of the proposed method increases as $c_1$ and $c_2$ become more distinct. }
\label{figure1_plot}
\end{figure}

\section{Discussion}

\setcounter{equation}{0}

{The FDR provides a practical and powerful approach to large-scale multiple testing problems and has been widely used in a wide range of scientific studies. The wFDR framework extends the FDR paradigm to integrate useful domain knowledge in simultaneous inference. The weights must be chosen with caution to avoid improper manipulation of results. }

In the multiple testing literature, procedural, decision, and class weights are often viewed as distinct weighting schemes and have been mostly investigated separately. Although this paper focuses on the decision weights approach,  the decision-theoretic framework enables a unified investigation of other weighting schemes. For example, a comparison of the LR (\ref{T_LR}) and WLR (\ref{T_IT}) demonstrates how the LR statistic may be adjusted optimally to account for the decision gains or losses. This shows that procedural weights may be derived in the decision weights framework. Moreover, the difference between the WLR (\ref{T_IT}) and WPO (\ref{T_PF}) shows the important role that $p_i$ plays in multiple testing. In particular the WPO (\ref{T_PF}) provides important insights on how prior beliefs may be incorporated in a decision weights approach to derive appropriate class weights. To see this, consider the multi-class model (\ref{multi-group}). Following the arguments in Cai and Sun (2009), we can conclude that in order to maximize the power, different FDR levels should be assigned to different classes. Similar suggestions for varied class weights have been made in Westfall and Young (1993, pages 169 and 186). These examples demonstrate that the decision weights approach provides a powerful framework to derive both procedural weights and class weights.

{Our formulation requires that the weights must be \emph{pre-specified} based on \emph{external} domain knowledge. It is of  interest to extend the work to the setting where the weights are unknown. Due to the variability in the quality of external information, subjectivity of investigators, and complexity in modeling and analysis, a systematic study of the issue is beyond the scope of the current paper. The optimal choice of weights depends on statistical, economic and scientific concerns jointly. Notable progresses have been made, for example, in Roeder and Wasserman (2009) and Roquain and van de Wiel (2009). However, these methods are mainly focused on the weighted $p$-value approach under the unweighted FDR criterion, hence do not apply to the framework in Benjamini and Hochberg (1997). Moreover, the optimal decision rule in the wFDR problem in general is not a thresholding rule based on the adjusted $p$-values. Much work is still needed to derive decision weights that would optimally incorporate domain knowledge in large-scale studies. }

\section{Proofs}\label{SecA}
\setcounter{equation}{0}

This section proves Theorem 1. The proofs of other results are given in the Supplementary Material.

\noindent \underline{\emph{Proof of  Part (i) of Theorem 1}}. To show that $\mbox{wFDR}(\pmb\delta_{OR}) = \alpha$,  we only need to establish that 
\begin{equation}
E_{U, \pmb a, \pmb b, \pmb X} \left\{ \sum_{i = 1}^m a_i \delta_{OR}^i(\mbox{Lfdr}_i - \alpha) \right\}= 0,\notag
\end{equation}
where the notation $E_{U, \pmb a, \pmb b, \pmb X}$ denotes that the expectation is taken over $U, \pmb a, \pmb b$, and $\pmb X$. According to the definitions of the capacity function $C(\cdot)$ and threshold $t^*$, we have 
\[
\sum_{i = 1}^m a_i \delta_{OR}^i(\mbox{Lfdr}_i-\alpha)=C(k)+I(U<p^*)\{C(k+1)-C(k)\}.
\]
It follows from the definition of $p^*$ that $
E_{U|\pmb a, \pmb b, \pmb X} \left \{ \sum_{i = 1}^m a_i \delta_{OR}^i (\mbox{Lfdr}_i - \alpha) \right \}= C(k) + \{C(k+1) - C(k)\} p^*=0,
$
where the notation $E_{U|\pmb a, \pmb b, \pmb X}$ indicates that the expectation is taken over $U$ while holding $(\pmb a, \pmb b, \pmb X)$ fixed. Therefore
\begin{equation}\label{a.3.1}
E_{U, \pmb a, \pmb b, \pmb X} \left\{ \sum_{i = 1}^m a_i \delta_{OR}^i(\mbox{Lfdr}_i - \alpha) \right\}= 0,
\end{equation}
and the desired result follows.

\noindent \underline{\emph{Proof of Part (ii) of Theorem 1}}. Let $\pmb \delta^*$ be an arbitrary decision rule such that $wFDR(\pmb \delta^*) \leq \alpha$. Then
\begin{equation} \label{a.3.2}
E_{\pmb a, \pmb b, \pmb X} \left\{\sum_{i = 1}^m a_i E(\delta_i^*|\pmb a, \pmb b, \pmb x)(\mbox{Lfdr}_i - \alpha)\right\} \leq 0.
\end{equation}
The notation $E(\delta_i^* |\pmb x, \pmb a, \pmb b)$ means that the expectation is taken to average over potential randomization conditional on the observations and weights. 

Let $\mathcal I^+=\{i: \delta_{OR}^i - E(\delta_i^* |\pmb x, \pmb a, \pmb b) > 0\}$ and $\mathcal I^-=\{i: \delta_{OR}^i - E(\delta_i^* |\pmb x, \pmb a, \pmb b) < 0\}$. For $i \in \mathcal I^+$, we have $\delta_{OR}^i = 1$ and hence $b_i(1-\mbox{Lfdr}_i) \geq t^* a_i(\mbox{Lfdr}_i - \alpha)$. Similarly for $i \in \mathcal I^-$, we have $\delta_{OR}^i = 0$ and so $b_i(1-\mbox{Lfdr}_i) \leq t^* a_i(\mbox{Lfdr}_i - \alpha)$. Thus
\[
\sum_{i \in \mathcal I^+ \cup \mathcal I^-} \left\{\delta_{OR}^i - E(\delta_i^* |\pmb x, \pmb a, \pmb b)\right\} \left\{b_i(1-\mbox{Lfdr}_i) - t^* a_i(\mbox{Lfdr}_i - \alpha)\right\} \geq 0. 
\]
Note that $\delta_{OR}^i$ is perfectly determined by $\pmb X$ except for $(k+1)$th decision. Meanwhile, $b_{(k+1)}\left(1-\mbox{Lfdr}_{(k+1)}\right) - t^* a_{(k+1)}\left(\mbox{Lfdr}_{(k+1)} - \alpha\right) = 0$ by our choice of $t^*$. It follows that
\begin{equation} \label{a.3.3}
E_{\pmb a, \pmb b, \pmb X}\left[\sum_{i = 1}^m \left\{E(\delta_{OR}^i|\pmb x, \pmb a, \pmb b) - E(\delta_i^* |\pmb x, \pmb a, \pmb b)\right\} \left\{b_i(1-\mbox{Lfdr}_i)- t^* a_i(\mbox{Lfdr}_i - \alpha)\right\}\right] \geq 0.
\end{equation}
Recall that the power function is given by $\mbox{ETP}(\pmb \delta)=E \left\{\sum_{i = 1}^m E(\delta_i|\pmb x, \pmb a, \pmb b)  b_i(1-\mbox{Lfdr}_i)\right\}$ for any decision rule $\pmb\delta$. Combining equations (\ref{a.3.1}) -- (\ref{a.3.3}) and noting that $t^* > 0$, we claim that $\mbox{ETP}(\pmb \delta_{OR}) \geq \mbox{ETP}(\pmb \delta^*) $  and the desired result follows. \qed

\newpage

\appendix

\section{Supplementary Theoretical and Numerical Results}

\setcounter{equation}{0}

%
%

This supplement contains additional discussions, technical analyses and numerical results. It is organized as follows. Section \ref{ranking.sec} discusses the optimal ranking issue and explains the difference between the VCR and WPO using a concrete example. The proofs of Theorem 2 and Propositions in the main paper are given in Sections \ref{Proof-Thm2} and \ref{Proof-Props}, respectively. Section \ref{equiv.sec} proves the equivalence of two wFDR criteria defined in Section 2.2 in the main paper. Additional numerical results are provided in Section \ref{Supp-Simulation.sec}. 

\appendix

\section{Optimal Ranking: VCR vs.~WPO}\label{ranking.sec}

In this section, we continue our discussion in Section 3.2 of the main paper regarding the optimal ranking in the wFDR formulation.

Although the WPO is optimal for PFER control, it is suboptimal for wFDR control. This section discusses a toy example to provide some insights on why the WPO ranking is dominated by the VCR ranking. We simulate 1000 $z$-values from a mixture model $(1-p) N(0,1)+ p N(2, 1)$ with $p=0.2$. The weights $a_i$ are fixed at 1 for all $i$, and $b_i$ are generated from log-normal distribution with location parameter $\ln 3$ and scale parameter 1. At wFDR level $\alpha=0.10$, we can reject 68 hypotheses along the WPO ranking, with the number of true positives being 60; in contrast, we can reject 81 hypotheses along the VCR ranking, with the number of true positives being 73. This shows that the VCR ranking enables us to ``pack more objects'' under the capacity $\mbox{wFDR}=0.1$ compared to the WPO ranking. Detailed simulation results are presented in Section E.2.

Next we give some intuitions on why the VCR ranking is more efficient in the wFDR problem. The test level $\alpha$, which can be viewed as the initial capacity for the error rate, plays an important role in the ranking process. Under the wFDR criterion, the capacity may either increase or decrease when a new rejection is made; the quantity that affects the current capacity is the \emph{excessive error rate} ($\mbox{Lfdr}_i - \alpha$). A different $\alpha$ would yield a different excessive error rate and hence a different ranking. (This is very different from the PFER criterion, under which the capacity always decreases when a new rejection is made and $\alpha$ is not useful in ranking.)  The next example shows that, although the WPO ranking always remains the same the VCR ranking can be altered by the choice of $\alpha$. 

\begin{example}{\rm
Consider two units $A$ and $B$ with observed values and weights $x_A=2.73$, $x_B=3.11$, $b_A=83.32$, and $b_B=11.95$. The Lfdr values are $\mbox{Lfdr}_A = 0.112$ and $\mbox{Lfdr}_B=0.055$, ranking $B$ ahead of $A$. Taking into account of the decision weights, the WPO values are $\mbox{WPO}_A=0.0015$ and $\mbox{WPO}_B=0.0049$, ranking $A$ ahead of $B$, and this ranking remains the same at all wFDR levels. At $\alpha=0.01$, we have $\mbox{VCR}_A=725.4$ and $\mbox{VCR}_B=250.9$, yielding the same ranking as the WPO. However, at $\alpha=0.05$, we have $\mbox{VCR}_A=1193.5$ and $\mbox{VCR}_B=2258.6$, reversing the previous ranking. This reversed ranking is due to the small excessive error rate $(\mbox{Lfdr}_B-\alpha)$ at $\alpha=0.05$, which makes the rejection of $B$, rather than $A$, more ``profitable''.
}
\end{example}

\section{Proof of Theorem 2}\label{Proof-Thm2}

\subsection{Notations and a useful lemma}\label{SecA.4}

We first recall and define a few useful notations. Let $\mathcal I_A$ be an indicator function, which equals 1 if event $A$ occurs and 0 otherwise. Let
$ N_i = a_i (\mbox{Lfdr}_i - \alpha)$,  $\widehat{N}_i = a_i (\widehat{\mbox{Lfdr}}_i - \alpha)$, $ R_i = \frac{a_i(\mbox{Lfdr}_i - \alpha)}{b_i(1 - \mbox{Lfdr}_i) + a_i|\mbox{Lfdr}_i - \alpha|}$ $\widehat{R}_i = \frac{a_i(\widehat{\mbox{Lfdr}}_i - \alpha)}{b_i(1 - \widehat{\mbox{Lfdr}}_i) + a_i|\widehat{\mbox{Lfdr}}_i - \alpha|},$ 
 $Q(t) = \frac{1}{m} \sum_{i=1}^m N_i \mathcal{I}_{R_i \leq t}$ and  $\widehat{Q}(t) = \frac{1}{m} \sum_{i=1}^m\widehat{N}_i \mathcal{I}_{\widehat{R}_i \leq t}$  for $t \in [0, 1]$ . 
%
Note that $Q(t)$ and $\widehat{Q}(t)$, the estimates for oracle and data driven capacities, are non-decreasing and right-continuous. We can further define
\begin{equation}\label{lambda-or}
\lambda_{OR} = \sup\{t \in [0,1]: Q(t) \leq 0\} \mbox{ and } \widehat{\lambda} = \sup\{t \in [0,1]: \widehat{Q}(t) \leq 0\}.
\end{equation}
Next we construct a continuous version of $Q(\cdot)$ for later technical developments. Specifically, for $0 \leq R_{(k)} < t \leq R_{(k+1)}$, let
$
Q^c(t) = \{1-r(t)\} Q\left(R_{(k)}\right) + r(t) Q\left(R_{(k+1)}\right),
$
where $c$ indicates ``continuous'' and $r(t) = (t-R_{(k)})/(R_{(k+1)}-R_{(k)})$. Let $R_{(m+1)} = 1$ and $N_{(m+1)} = 1$. Similarly we can define a continuous version of $\hQ(t)$. For $0 \leq \hR_{(k)} < t \leq \hR_{(k+1)}$, let
$
\hQ^c(t) = [1-\hr(t)] \hQ(\hR_{(k)}) + \hr(t) \hQ(\hR_{(k+1)}),
$
with $\hr(t) = (t-\hR_{(k)})/(\hR_{(k+1)}-\hR_{(k)})$.
Now the inverses of $Q^c(t)$ and $\hQ^c(t)$ are well defined; denote these inverses by $Q^{c,-1}(t)$ and $\hQ^{c,-1}(t)$, respectively. By construction, it is easy to see that
\begin{equation} \label{result_from_notation}
\mathcal{I}_{R_i \leq \lambda_{OR}} = \mathcal{I}_{R_i \leq Q^{c, -1}(0)} \mbox{ and }\mathcal{I}_{\hR_i \leq \hlambda} = \mathcal{I}_{\hR_i \leq \hQ^{c, -1}(0)}.
\end{equation}

Next we  state and prove a lemma that contains some key facts to prove the theorem. 
\begin{lemma}\label{basic} Assume that Conditions 1-3 hold. For any $t \in [0,1]$, we have 
\begin{description}
 \item (i) $E\left(\hN_i \mathcal{I}_{[\hR_i \leq t]}-N_i\mathcal{I}_{[R_i \leq t]}\right)^2= o(1)$,
 \item (ii) $E\left\{\left(\hN_i \mathcal{I}_{[\hR_i \leq t]}-N_i\mathcal{I}_{[R_i \leq t]}\right)\left(\hN_j \mathcal{I}_{[\hR_j \leq t]}-N_j\mathcal{I}_{[R_j \leq t]}\right)\right\}= o(1)$, and 
 \item (iii) $\hQ^{c, -1}(0) - Q^{c, -1}(0) \convp 0$.
\end{description}
\end{lemma}

\subsection{Proof of Lemma}

\noindent\underline{\emph{Proof of Part (i)}}. 
We first decompose $E\left(\hN_i \mathcal{I}_{[\hR_i \leq t]}-N_i\mathcal{I}_{[R_i \leq t]}\right)^2$ into three terms:
\begin{eqnarray}\label{decomp}
&& E\left(\hN_i \mathcal{I}_{[\hR_i \leq t]}-N_i\mathcal{I}_{[R_i \leq t]}\right)^2 \nonumber \\ & = & E[(\hN_i -N_i)^2\mathcal{I}_{\hR_i \leq t, R_i \leq t}] + E[\hN_i^2 \mathcal{I}_{\hR_i \leq t, R_i > t}]
+ E[N_i^2 \mathcal{I}_{\hR_i > t, R_i \leq t}].
\end{eqnarray}
Next we argue below that all three terms are of $o(1)$.


First, it follows from the definitions of $\hN_i$ and $N_i$ that
\begin{align*}
E\left\{(\hN_i -N_i)^2\mathcal{I}_{\hR_i \leq t, R_i \leq t}\right\}  &= E\left\{a_i^2 \left(\mbox{Lfdr}_i - \widehat{\mbox{Lfdr}}_i\right)^2\mathcal{I}_{\hR_i \leq t, R_i \leq t}\right\} \\
&\leq E\left\{a_i^2 \left(\mbox{Lfdr}_i - \widehat{\mbox{Lfdr}}_i\right)^2\right\}.
\end{align*}
By an application of Cauchy-Schwarz inequality, we have
\[
E\left\{a_i^2 \left(\mbox{Lfdr}_i - \widehat{\mbox{Lfdr}}_i\right)^2\right\} \leq \left\{E(a_i^4)\right\}^{1/2} \left\{E\left(\mbox{Lfdr}_i - \widehat{\mbox{Lfdr}}_i\right)^4\right\}^{1/2}.
\]
It follows from Condition 2 that $E(a_i^4) = O(1)$. To show $E\left(\mbox{Lfdr}_i - \widehat{\mbox{Lfdr}}_i\right)^4 = o(1)$, note that both $\mbox{Lfdr}_i$ and $\widehat{\mbox{Lfdr}}_i$ are in $[0,1]$. Hence $E\left(\mbox{Lfdr}_i - \widehat{\mbox{Lfdr}}_i\right)^4 \leq E|\mbox{Lfdr}_i - \widehat{\mbox{Lfdr}}_i|$. Using the fact that $\mbox{Lfdr}_i - \widehat{\mbox{Lfdr}}_i = o_P(1)$, the uniform integrability for bounded random variables, and the Vitali convergence theorem,  we conclude that $E|\mbox{Lfdr}_i - \widehat{\mbox{Lfdr}}_i| = o(1)$. Therefore, the first term in (\ref{decomp}) is of $o(1)$.

Next we show that $E\left(\hN_i^2 \mathcal{I}_{\hR_i \leq t, R_i > t}\right) = o(1)$. Applying Cauchy-Schwarz inequality again, we have
$
E\left(\hN_i^2 \mathcal{I}_{\hR_i \leq t, R_i > t}\right) \leq (1-\alpha)^2\left\{E(a_i^4)\right\}^{1/2} \left\{P\left(\hR_i \leq t, R_i > t\right)\right\}^{1/2}.
$
Condition 2 implies that $E(a_i^4) = O(1)$; hence we only need to show that $P(\hR_i \leq t, R_i > t) = o(1)$. 
Let $\eta>0$ be a small constant. Then 
\begin{align*}
P(\hR_i \leq t, R_i > t) &= P\left(\hR_i \leq t, R_i \in (t, t+\eta]\right) + P\left(\hR_i \leq t, R_i > t + \eta\right) \\
&\leq P(R_i \in (t, t+\eta]) + P(|\hR_i - R_i| > \eta).
\end{align*}
Since $R_i$ is a continuous random variable, we can find $\eta_t > 0$ such that $P(R_i \in (t, t+\eta]) < \varepsilon/2$ for a given $\varepsilon$. For this fixed $\eta_t >0$, we can show that $P(|\hR_i - R_i| > \eta_t) < \varepsilon/2$ for sufficiently large $n$. This follows from $\mbox{Lfdr}_i - \widehat{\mbox{Lfdr}}_i = o_P(1)$ and the continuous mapping theorem. Similar argument can be used to prove that $E[N_i^2 \mathcal{I}_{\hR_i > t, R_i \leq t}] = o(1)$, hence completing the proof of part (i).
\smallskip

\noindent\underline{\emph{Proof of Part (ii)}}. As $X_i$ and $X_j$ are identically distributed and our estimates are invariant to permutation, we have
\[
E\left\{(\hN_i \mathcal{I}_{[\hR_i \leq t]}-N_i\mathcal{I}_{[R_i \leq t]})(\hN_j \mathcal{I}_{[\hR_j \leq t]}-N_j\mathcal{I}_{[R_j \leq t]})\right\}  \leq E\left(\hN_i \mathcal{I}_{[\hR_i \leq t]}-N_i\mathcal{I}_{[R_i \leq t]}\right)^2.
\]
The desired result follows from part (i). 

\smallskip
\noindent\underline{\emph{Proof of Part (iii)}}.  Define $Q_{\infty}(t) = E(N_i \mathcal{I}_{R_i \leq t})$, where the expectation is taken over $(\pmb a, \pmb b, \pmb X, \pmb\theta)$. Let
\[
\lambda_{\infty} = \sup\{t \in [0,1]: Q_{\infty}(t) \leq 0\}. 
\]
We will show that (i) $Q^{c, -1}(0) \convp \lambda_{\infty}$ and  (ii) $\hQ^{c, -1}(0) \convp \lambda_{\infty}$. Then the desired result $\hQ^{c, -1}(0) - Q^{c, -1}(0) \convp 0$ follows from (i) and (ii).

Fix $t \in [0,1]$. By Condition 2 and WLLN, we have that $Q(t) \convp Q_{\infty}(t)$. Since $Q^{c, -1}(\cdot)$ is continuous, for any $\varepsilon > 0$, there exists a $\delta > 0$ such that $|Q^{c, -1}(Q_\infty(\lambda_{\infty})) - Q^{c, -1}(Q^c(\lambda_{\infty}))| < \varepsilon$ whenever $|Q_\infty(\lambda_{\infty}) -Q^c(\lambda_{\infty})| < \delta$. It follows that
\begin{eqnarray}\label{eq:Q-c1}
&& P\left\{|Q_\infty(\lambda_{\infty}) -Q^c(\lambda_{\infty})| > \delta\right\}  \\ & \geq &  P\left\{|Q^{c, -1}(Q_\infty(\lambda_{\infty})) - Q^{c, -1}(Q^c(\lambda_{\infty}))| > \varepsilon\right\} \nonumber \\ \label{eq:Q-c}
& = & P\left\{|Q^{c, -1}(0) - \lambda_{\infty}| > \varepsilon\right\}. 
\end{eqnarray}
Equation \eqref{eq:Q-c} holds since $Q_\infty(\lambda_{\infty})=0$ by the continuity of $R_i$, and $Q^{c, -1}(Q^c(\lambda_{\infty})) = \lambda_{\infty}$ by the definition of inverse. Therefore we only need to show that for any $t \in [0,1], Q^c(t) \convp Q_{\infty}(t)$. Note that $
E|Q(t) - Q^c(t)| \leq \frac{E(\sup_i a_i)}{m} \rightarrow 0,
$
by Condition 2. Using Markov's inequality, $Q(t) - Q^c(t) \convp 0$. Following from $Q(t) \convp Q_{\infty}(t)$, we have $Q^c(t) \convp Q_{\infty}(t)$. Therefore \eqref{eq:Q-c1} and hence \eqref{eq:Q-c} goes to 0 as $m\rightarrow\infty$, establishing the desired result (i) $Q^{c, -1}(0) \convp \lambda_{\infty}$.   

To show result (ii) $\hQ^{c, -1}(0) \convp \lambda_{\infty}$, we can repeat the same steps. In showing $Q^{c, -1}(0) \convp \lambda_{\infty}$, we only used the facts that (a) $Q(t) \convp Q_{\infty}(t)$, (b) $Q^{c, -1}(\cdot)$ is continuous, and (c) $Q(t) - Q^c(t) \convp 0$. Therefore to prove $\hQ^{c, -1}(0) \convp \lambda_{\infty}$, we only need to check whether the same conditions (a) $\hQ(t) \convp Q_{\infty}(t)$, (b) $\hQ^{c, -1}(\cdot)$ is continuous, and (c) $\hQ(t) - \hQ^c(t) \convp 0$ still hold. It is easy to see that (b) holds by definition, and (c) holds by noting that
\[
E|\hQ(t) - \hQ^c(t)| \leq \frac{E(\sup_i a_i)}{m} \rightarrow 0.
\]
The only additional result we need to establish is (a).

Previously, we have shown  that $Q(t)\convp Q_\infty(t)$. Therefore the only additional fact that we need to establish is that $|\hQ(t) - Q(t)|\convp 0$. Now consider the following quantity:
\begin{equation}\label{Delta-Q}
\Delta Q=\{\hQ(t)-Q(t)\}-[E\{\hQ(t)\}-E\{Q(t)\}].
\end{equation}
By repeating the steps of part (i) we can show that 
\begin{equation}\label{hQ-0}
|E\{\hQ(t)\}-E\{Q(t)\}|=|E(N_i\mathcal{I}_{R_i \leq t}) - E(\hN_i \mathcal{I}_{\hR_i \leq t})| \rightarrow 0.
\end{equation} 
By definition, $\Delta Q={m}^{-1} \sum_{i=1}^m\{\hN_i \mathcal{I}_{[\hR_i \leq t]}-N_i\mathcal{I}_{[R_i \leq t]}\} - [E(\hN_i \mathcal{I}_{\hR_i \leq t})-E(N_i \mathcal{I}_{R_i \leq t})].$ For an application of WLLN for triangular arrays (see, for e.g., Theorem 6.2 of Billingsley, 1991), we need to show that $\text{var}(\sum_{i=1}^m\{\hN_i \mathcal{I}_{[\hR_i \leq t]}-N_i\mathcal{I}_{[R_i \leq t]}\})/m^2 \rightarrow 0.$
Using the result in Part (i) we deduce that,
\begin{eqnarray*}
&& m^{-2}\text{Var}\left\{\sum_{i=1}^m\left(\hN_i \mathcal{I}_{[\hR_i \leq t]}-N_i\mathcal{I}_{[R_i \leq t]}\right)\right\} \leq m^{-2}E\left\{\sum_{i=1}^m\left(\hN_i \mathcal{I}_{[\hR_i \leq t]}-N_i\mathcal{I}_{[R_i \leq t]}\right)\right\}^2 \\
&= & \left(1-\frac{1}{m}\right)E\left\{\left(\hN_i \mathcal{I}_{[\hR_i \leq t]}-N_i\mathcal{I}_{[R_i \leq t]}\right)\left(\hN_j \mathcal{I}_{[\hR_j \leq t]}-N_j\mathcal{I}_{[R_j \leq t]}\right)\right\} \\ && +\frac{1}{m}E\left(\hN_i \mathcal{I}_{[\hR_i \leq t]}-N_i\mathcal{I}_{[R_i \leq t]}\right)^2
= o(1).
\end{eqnarray*}
It follows from the WLLN for triangular arrays that $|\Delta Q|\convp 0$. 
Combining (\ref{Delta-Q}) and (\ref{hQ-0}), we conclude that $|\hQ(t) - Q(t)|\convp 0$,
which completes the proof.
\qed

\subsection{Proof of Theorem 2}

\noindent\underline{\emph{Proof of Part (i)}}.
Consider the oracle and data driven thresholds $\lambda_{OR}$ and $\hlambda$ defined in Equation \eqref{lambda-or}. The wFDRs of the oracle and data-driven procedures are 
$$\mbox{wFDR}_{OR}  =  \frac{E\left\{\sum\limits_i a_i (1-\theta_i)\delta_{OR}^i\right\}}{E\left(\sum\limits_i a_i \delta_{OR}^i\right)},  \quad
\mbox{wFDR}_{DD}  =  \frac{E\left\{\sum\limits_i a_i (1-\theta_i)\mathcal{I}_{\widehat{R}_i \leq \hlambda}\right\}}{E\left(\sum\limits_i a_i \mathcal{I}_{\widehat{R}_i \leq \hlambda}\right)}.
$$Making the randomization explicit, the wFDR of the oracle procedure is
\[
\mbox{wFDR}_{OR} = \frac{E\left\{m^{-1}\sum\limits_i a_i (1-\theta_i)\mathcal{I}_{R_i \leq \lambda_{OR}}+m^{-1}a_{i^*}(1-\theta_{i^*})\delta_{OR}^{i^*}\right\}}{E\left(m^{-1}\sum\limits_i a_i \mathcal{I}_{R_i \leq \lambda_{OR}}+m^{-1}a_{i^*}\delta_{OR}^{i^*}\right)},
\]
where $i^*$ indicates the randomization point in a realization.  Note that both $E\{a_{i^*}(1-\theta_{i^*})\delta_{OR}^{i^*}/m\}$ and $E\{a_{i^*}\delta_{OR}^{i^*}/m\}$ are bounded by $E(a_{i^*}/m)$. Hence by Condition 2 both quantities are of $o(1)$. 


From the discussion in Section B.2, $\mathcal{I}_{R_i \leq \lambda_{OR}} = \mathcal{I}_{R_i \leq Q^{c, -1}(0)}$ and $\mathcal{I}_{\hR_i \leq \hlambda} = \mathcal{I}_{\hR_i \leq \hQ^{c, -1}(0)}$. According to Part (iii) of Lemma \ref{basic}, we have $\{\hR_i - \hQ^{c, -1}(0)\} - \{R_i - Q^{c, -1}(0)\} = o_P(1)$. Following the proof of Lemma \ref{basic} that
$$
E \left\{a_i (1-\theta_i)\mathcal{I}_{\hR_i - \hQ^{c, -1}(0) \leq 0}\right\} = E \left\{a_i (1-\theta_i)\mathcal{I}_{R_i - Q^{c, -1}(0) \leq 0}\right\} + o(1).
$$
It follows that $m^{-1} E \{\sum_i a_i (1-\theta_i)\mathcal{I}_{\hR_i \leq \hlambda}\} \rightarrow m^{-1} E \{\sum_i a_i (1-\theta_i)\mathcal{I}_{R_i \leq \lambda_{OR}}\}. $ 
Similarly, we can show that
$E \left(a_i\mathcal{I}_{\hR_i - \hQ^{c, -1}(0) \leq 0}\right) = E \left(a_i\mathcal{I}_{R_i - Q^{c, -1}(0) \leq 0}\right) + o(1). 
$
Further from Condition 2 the quantity $m^{-1} E (\sum_i a_i \mathcal{I}_{R_i \leq \lambda_{OR}})$ is bounded away from zero. To see this, note that Condition 1 implies that $a_i$ is independent of $\mbox{Lfdr}_i$. It follows that 
$
m^{-1} E \left(\sum_{i = 1}^m a_i \mathcal{I}_{R_i \leq \lambda_{OR}}\right)= E \left(a_i \mathcal{I}_{R_i \leq \lambda_{OR}}\right) \geq c\widetilde{p}_{\alpha} > 0,
$
where $P(\mbox{Lfdr}(X) \leq \alpha) \geq \widetilde{p}_{\alpha}$ for some $\widetilde{p}_{\alpha} \in (0,1]$ for the choice of the nominal level $\alpha \in (0,1)$ and $X$, an i.i.d copy of $X_i$. This holds for any non-vanishing $\alpha$. (Note that all hypotheses with $\mbox{Lfdr}_i<\alpha$ will be rejected automatically).
Therefore we conclude that $\mbox{wFDR}_{DD} = \mbox{wFDR}_{OR} + o(1) = \alpha + o(1).$

\smallskip
\noindent\underline{\emph{Proof of Part (ii)}}. The quantity $m^{-1}ETP_{DD}$ is defined as $m^{-1}E\left(b_i\theta_i\mathcal{I}_{[\hR_i \leq \hlambda]}\right)$. Making the randomization explicit, we have
$
m^{-1}ETP_{OR} = E\left(\frac{1}{m}\sum_i b_i \theta_i \mathcal{I}_{[R_i \leq \lambda_{OR}]}+\frac{1}{m}b_{i^*}\theta_{i^*}\delta_{OR}^{i^*}\right),
$
where $i^*$ indicates the randomized point. By Condition 2, 
$
0 \leq m^{-1}E\left(b_{i^*}\theta_{i^*}\delta_{OR}^{i^*}\right) \leq \frac{Eb_{i^*}}{m} \leq \frac{E \sup_i b_{i}}{m} = o(1).
$
From the discussion in Section B.2, $\mathcal{I}_{R_i \leq \lambda_{OR}} = \mathcal{I}_{R_i \leq Q^{c, -1}(0)}$ and $\mathcal{I}_{\hR_i \leq \hlambda} = \mathcal{I}_{\hR_i \leq \hQ^{c, -1}(0)}$. Repeating the steps in proving the wFDR, we can show that
$E\left(b_i \theta_i \mathcal{I}_{[\hat R_i \leq \hat\lambda]} \right)=E\left(b_i \theta_i \mathcal{I}_{[R_i \leq \lambda_{OR}]} \right)+o(1). $
Finally, it is easy to show that 
$
E\left(b_i \theta_i \mathcal{I}_{[R_i \leq \lambda_{OR}]} \right)
\geq c(1-\alpha)\widetilde{p}_\alpha,
$
which is bounded below by a nonzero constant. Here the positive constant $c$ is as defined in Condition 2. We conclude that
$ETP_{DD}/ETP_{OR} = 1 + o(1)$.  \qed

\section{Proofs of Propositions in the Main Paper}
 \label{Proof-Props}

\subsection{Proof of Proposition 2}

Let $\lambda > 0$ be the relative cost of a false positive to a false negative. Consider the following weighted classification problem with loss function: 
\begin{equation}\label{WCL-problem}
 L_{\pmb a, \pmb b}(\pmb\theta, \pmb\delta)= \sum_{i=1}^m \{ \lambda a_i (1-\theta_i)\delta_i+ b_i\theta_i(1-\delta_i)\}.
\end{equation}
We aim to find $\pmb\delta$ that minimizes the posterior loss $E_{\pmb\theta|\pmb X} \{L_{\pmb a, \pmb b}(\pmb\theta, \pmb\delta)\}$
\begin{align}
&\arg\!\min_{\pmb\delta} \sum_{i=1}^m \left\{ \lambda a_i P(\theta_i = 0| X_i) \delta_i+ b_i P(\theta_i = 1|X_i)(1-\delta_i)\right\} \nonumber \\
= &\arg\!\min_{\pmb\delta} \sum_{i=1}^m \left\{ \lambda a_i P(\theta_i = 0|X_i) - b_i P(\theta_i = 1|X_i)\right\} \delta_i \nonumber.
\end{align}
Therefore the optimal decision rule $\pmb\delta_{PF}=(\delta_{PF}^i: i=1, \cdots, m)$ is of the form
\begin{equation}
\delta_{PF}^i = I\left[\frac{a_iP(\theta_i = 0|X_i)}{b_iP(\theta_i = 1|X_i)} < \frac{1}{\lambda}\right],
\end{equation}
which reduces to the test statistic defined in (3.14).

Next note that $Q_{PF}(t)$ is a continuous and increasing function of $t$. Therefore we can find $t_{PF}$ such that $Q_{PF}(t_{PF})=\alpha$. 
For an arbitrary decision rule $\pmb\delta^* \in \mathcal D_\alpha$, we must have $ETP(\pmb\delta^*) \leq ETP(\pmb\delta_{PF})$. Suppose not, then there exists $\pmb\delta^* \in \mathcal D_\alpha$ such that $\mbox{PFER} (\pmb\delta^*) \leq \alpha = \mbox{PFER}(\pmb\delta_{PF})$ and $-\mbox{ETP}(\pmb\delta^*) < -\mbox{ETP}(\pmb\delta_{PF})$. Consider a weighted classification problem with $\lambda = 1/t_{PF}$. Then we can show that $\pmb\delta^*$ has a smaller classification risk compared to $\pmb\delta_{PF}$. This is a contradiction. Therefore we must have $\mbox{ETP}(\pmb\delta^*)\leq \mbox{ETP}(\pmb\delta_{PF})$. \qed

\subsection{Proof of Proposition 3} \label{SecA.2}

\noindent\underline{\emph{Proof of Part (i)}}. For convenience of notation, define $S_i = 1/\mbox{VCR}_i$. We show that rankings by increasing values of $R_i$ and $S_i$ are the same. If $i\in S^+$, then all values are positive. Sorting by increasing $S_i$ is the same as sorting by decreasing $(1/S_i) + 1$ and hence by increasing $1/(1/S_i + 1)$, which is precisely sorting by increasing $R_i$. If $i\in S^-$, then all values are negative. Sorting by increasing $S_i$ is the same as sorting by decreasing $(1/S_i) - 1$  and hence by increasing $1/(1/S_i - 1)$, which is again the same as sorting by increasing $R_i$. The desired result follows.

\noindent\underline{\emph{Proof of Part (ii)}}. The result follows directly from the facts that (a) $R_i$ is negative when $i\in S^-$ and (b) $R_i$ is positive if $i\in S^+$. \qed

\section{Asymptotic Equivalence of the wFDR Definitions}\label{equiv.sec}
To establish the 
asymptotic equivalence, we first give a sufficient condition (Proposition 1) and then show that the condition is fulfilled by our proposed data-driven procedure (Proposition \ref{weight_eqv_dd}). Simulation results are provided in Section E.3 to compare the two definitions in finite samples. 
\begin{proposition}\label{lemma_weight_eqv}
Consider a general decision rule $\pmb \delta$. Let $\mathcal{Y} = m^{-1} \sum_{i = 1}^m a_i \delta_i $.  Then
 $\mbox{wFDR}(\pmb\delta) = \mbox{wFDR}_{\rm{BH}}(\pmb\delta) + \text{o}(1)$ if 
\begin{equation}\label{equi-cond}
 (a) \mbox{$E\mathcal{Y}\geq \underline{\eta}$ for some $ \underline{\eta}>0$, and (b) $\mathrm{Var} \mathcal{Y} = \text{o}(1)$.} 
 \end{equation}
\end{proposition}

\noindent \textbf{Proof of Proposition \ref{lemma_weight_eqv}.} Let $\mathcal{X} = m^{-1}\sum\limits_{i=1}^m a_i (1-\theta_i)\delta_i$. Note that when $\mathcal{Y} = 0$ we must have $\mathcal{X} = 0$. The asymptotic equivalence follows if we can show the following
\begin{equation} \label{wBH_eq1}
\mbox{wFDR}(\pmb\delta) -\mbox{WFDR}_{\rm{BH}}(\pmb\delta) \leq E \left \{ \left | Q(\bf a) - \frac{\mathcal{X}}{E\mathcal{Y}} \right | \right \} = E\left \{ \left | \frac{\mathcal{X}}{\mathcal{Y}} - \frac{\mathcal{X}}{E\mathcal{Y}} \right | \mathcal{I}_{\mathcal{Y} > 0} \right \} = o(1).
\end{equation}
 Since $\mathcal{X} \leq \mathcal{Y}$ and both are non-negative expressions, using Cauchy-Schwarz 
\begin{equation} \label{wBH_eq2}
E \left \{ \left | \frac{\mathcal{X}}{\mathcal{Y}} - \frac{\mathcal{X}}{E\mathcal{Y}} \right | \mathcal{I}_{\mathcal{Y} > 0} \right \} = E \left \{ \frac{\mathcal{X}}{\mathcal{Y}} \mathcal{I}_{\mathcal{Y} > 0} \frac{\left |\mathcal{Y} - E \mathcal{Y} \right |}{E\mathcal{Y}} \right \} \leq \frac{(E \left |\mathcal{Y} - E \mathcal{Y} \right |^2)^{1/2}}{E\mathcal{Y}} = \frac {(\mathrm{Var} \mathcal{Y})^{1/2}} {E\mathcal{Y}}.
\end{equation}
Combining (\ref{wBH_eq2}), and Conditions (a) and (b), we establish (\ref{wBH_eq1}). \qed

The next proposition shows that the two weighted FDR definitions [(2.4) and (2.5)] are asymptotically equivalent when Procedure 1 is used.  
\begin{proposition}\label{weight_eqv_dd}
Consider random mixture model (2.1). Suppose Conditions 1-3 hold. Then condition (\ref{equi-cond}) is fulfilled by the data-driven procedure  $\pmb\delta_{\rm dd}$. Hence Procedure 1 controls  $\mbox{wFDR}_{\rm BH}$ defined in (2.4) at level $\alpha+o(1)$. 
\end{proposition}

\noindent\textbf{Proof of Proposition \ref{weight_eqv_dd}}. From proof of Theorem 2, part (i) we have
\begin{eqnarray*}
E \left(a_i\mathcal{I}_{\hR_i - \hQ^{c, -1}(0) \leq 0}\right) = E \left(a_i\mathcal{I}_{R_i - Q^{c, -1}(0) \leq 0}\right) + o(1), \\
m^{-1} E \left(\sum_{i = 1}^m a_i \mathcal{I}_{R_i \leq \lambda_{OR}}\right) =  E \left(a_i \mathcal{I}_{R_i \leq \lambda_{OR}}\right) \geq c\widetilde{p}_{\alpha} > 0.
\end{eqnarray*}
Let $\underline{\eta} = c\widetilde{p}_{\alpha}/2$ and $\itY' = m^{-1} \sumim a_i \itI_{R_i \leq \lambda_{\infty}}$. Note that 
\[
\mathrm{Var}  \itY' = m^{-1} \mathrm{Var} (a_i \itI_{R_i \leq \lambda_{\infty}}) \leq m^{-1} E(a_i^2) \leq m^{-1}(C+1) = o(1),
\] where the last inequality follows from Condition 2. To show that $\mathrm{Var} \mathcal{Y} = o(1)$, decompose
\[
\mathrm{Var} \mathcal{Y} = \mathrm{Var} \itY' + \mathrm{Var} (\itY - \itY') + 2\mathrm{Cov} (\itY - \itY', \itY'). 
\]
We only need to show that $\mathrm{Var} (\itY - \itY') = o(1)$. Then by Cauchy-Schwarz inequality and using $\mathrm{Var} \itY'  = o(1)$ it follows that $2\mathrm{Cov} (\itY - \itY', \itY') = o(1)$.

Recall that $\hR_i - R_i  = o_P(1)$ and $\hQ^{c, -1}(0) - \lambda_{\infty} = o_P(1)$. We have
\begin{eqnarray*}
&& m^{-2}\text{Var}\left\{\sum_{i=1}^m\left(a_i \mathcal{I}_{\hR_i - \hQ^{c, -1}(0) \leq 0}-a_i\mathcal{I}_{R_i - \lambda_{\infty} \leq 0}\right)\right\} \\
&\leq& m^{-2}E\left\{\sum_{i=1}^m\left(a_i \mathcal{I}_{\hR_i - \hQ^{c, -1}(0) \leq 0}-a_i\mathcal{I}_{R_i - \lambda_{\infty} \leq 0}\right)\right\}^2 \\
&= & \left(1-\frac{1}{m}\right)E\left\{\left(a_i \mathcal{I}_{\hR_i - \hQ^{c, -1}(0) \leq 0}-a_i\mathcal{I}_{R_i - \lambda_{\infty} \leq 0}\right)\left(a_j \mathcal{I}_{\hR_j - \hQ^{c, -1}(0) \leq 0}-a_j\mathcal{I}_{R_j - \lambda_{\infty} \leq 0}\right)\right\} \\ && +\frac{1}{m}E\left(a_i \mathcal{I}_{\hR_i - \hQ^{c, -1}(0) \leq 0}-a_i\mathcal{I}_{R_i - \lambda_{\infty} \leq 0}\right)^2
= o(1),
\end{eqnarray*}
where the last equality can be shown by following similar arguments in the proof of parts (i) and (ii) in Lemma 1. \qed

\section{Additional Numerical Results}\label{Supp-Simulation.sec}

In this section we provide the simulation results for general weights, as well as additional numerical evidence to support the claims in the main text.

\subsection{Additional results with group-wise weights}

In the second simulation study, we investigate the effect of the signal strength $\mu$. Similar as before, consider two groups of hypotheses with group sizes $m_1 = 3000$ and $m_2 = 1500$.  Under this setting $c_1$ and $c_2$ are fixed at 3 and 0.33, respectively. The non-null proportion is $p=0.2$ and the signal strength $\mu$ is varied from 1.75 to 2.5.  We apply different methods to the simulated data sets and obtain the wFDR and ETP levels as functions of $\mu$ by averaging results over 200 replications.  The simulation results are summarized in Figure \ref{figure2_plot}. We can see from Panel (a) that all methods control the wFDR at the nominal level 0.1 approximately (the BH95 method is very conservative and the result is not displayed). Panel (b) shows that the proposed methods dominates other competing methods; and the gain in power is more pronounced when the signals are weak. (The ETP increases rapidly with increased signal strength. For better visualization of results, we present the graph in a logarithmic scale. {See Table \ref{group-simu.tab} for results of the BH95 method, as well as the ETP levels in original scales.})

\begin{figure}
\centering
\includegraphics[max size={\textwidth}{\textheight}]{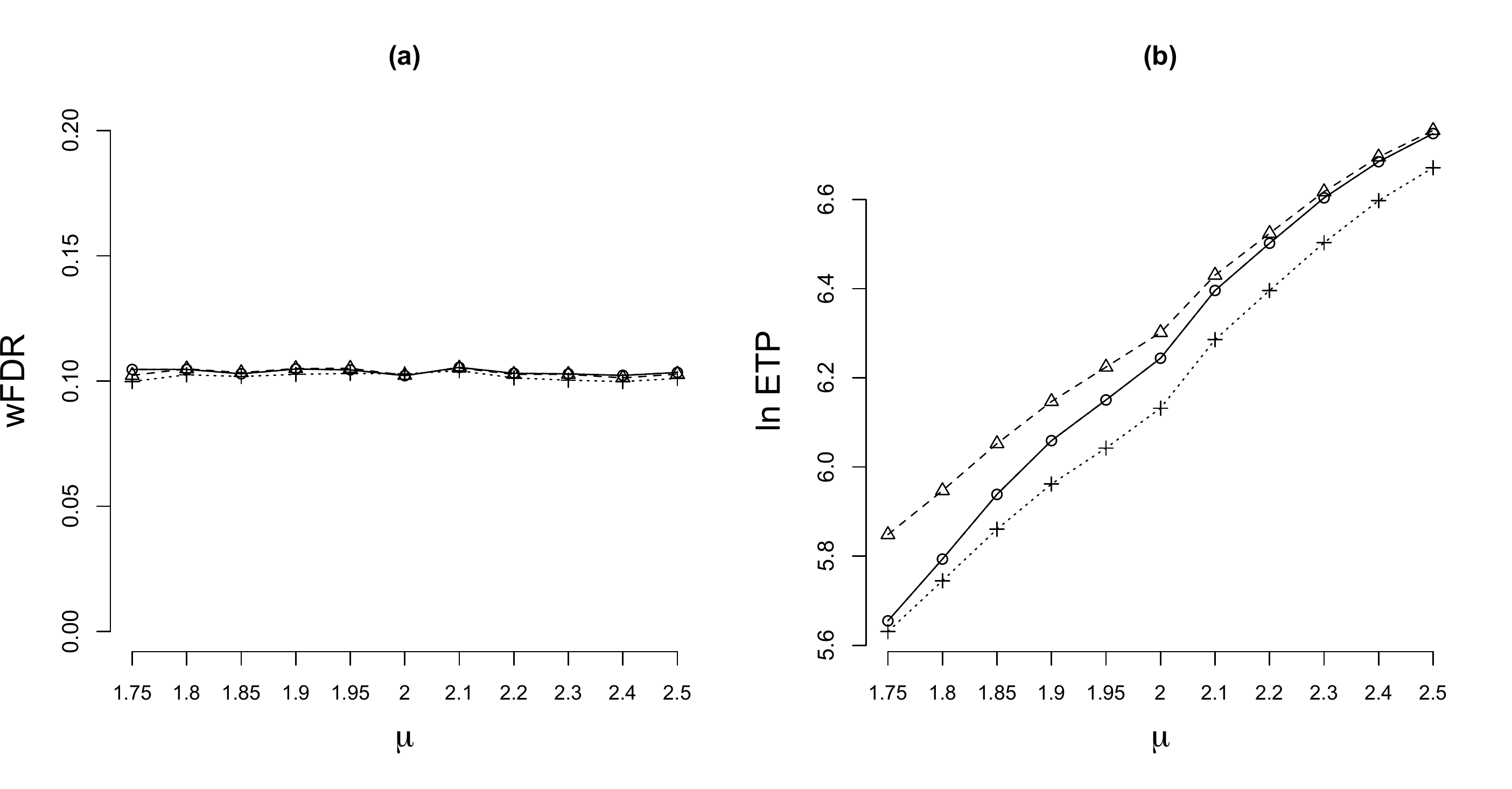} 
\caption{Comparison under group-wise weights: $\circ$ WPO, $\triangle$ DD (proposed), and $+$ AZ. The efficiency gain of the proposed method is more pronounced when signals are weak.}\label{group-simu.tab}
\label{figure2_plot}
\end{figure}

\subsection{General weights}\label{simulation3_sec}

In applications where domain knowledge is precise (e.g.~spatial cluster analysis), dividing the hypotheses into groups and assigning group-wise weights would not be satisfying. This section investigates the performance of our method when random weights $(a_i, b_i)$ are generated from a bivariate distribution. 

In the third simulation study, we test $m = 3000$ hypotheses with $a_i$, the weights associated with the wFDR control, fixed at 1. We generate $b_i$, the weights associated with the power (or ETP), from log-normal distribution with location parameter $\ln 3$ and scale parameter 1. The location parameter is chosen in a way such that the median weight is 3, similar to those in previous settings. We apply different methods with 200 replications. 

The simulation results are summarized in Figure \ref{figure3_plot}. The first row fixes $\alpha = 0.10$ and $p = 0.2$, and plots the wFDR and ETP as functions of $\mu$. The second row fixes  $\alpha = 0.10$ and $\mu=1.9$, and plots the wFDR and ETP as functions of $p$. The last row fixes $p=0.2$ and $\mu=1.9$, and plots the wFDR and ETP as functions of $\alpha$. In the plots, we omit the BH95 method (which is very conservative) and present the ETP in a logarithmic scale (for better visualization of results). The following observations can be made: (i) all methods control the wFDR at the nominal level approximately; (ii) by exploiting the weights $b_i$, the WPO method outperforms the unweighted AZ method; (iii) the proposed method outperforms all competing methods; (iv) Panel (f) shows that gains in power of the proposed method over the WPO method vary at different nominal levels $\alpha$; (v) similar to the observations in previous simulation studies, the difference between the WPO method and the proposed method decreases with increased signal strength, the efficiency gain of the proposed method is larger as signals become more sparse.

\begin{table}[h!]\caption{ETP values (in original scale) of various methods corresponding to Figure 
\
\ref{figure2_plot}}
\begin{center}
\resizebox{\textwidth}{!} {\begin{tabular}{c c c c c c c c c c c c} 
 \hline\hline
$\mu =$ & 1.75 & 1.80 & 1.85 & 1.90 & 1.95 & 2.0 & 2.1 & 2.2 & 2.3 & 2.4 & 2.5\\
 \hline
 BH95 & 102.5 & 125.8 & 150.6 & 179.6 & 204.6 & 237.0 & 301.7 & 361.5 & 431.2 & 501.0 & 567.4\\
 AZ & 278.9 & 312.6 & 350.9 & 388.3 & 420.9 & 460.4 & 536.8 & 599.4 & 667.2 & 733.2 & 789.4\\
 WPO & 285.7 & 328.1 & 379.4 & 428.0 & 468.9 & 514.9 & 599.4 & 666.4 & 737.8 & 800.1 & 852.3\\
 DD (proposed) & 346.7 & 382.6 & 425.1 & 467.3 & 504.8 & 545.4 & 620.4 & 681.3 & 748.1 & 808.7 & 858.2\\
 \hline
\end{tabular}}
\end{center}
\label{etp_table_1}
\end{table}


\begin{figure}
\centering
\includegraphics[width=\textwidth, height=2.6in]{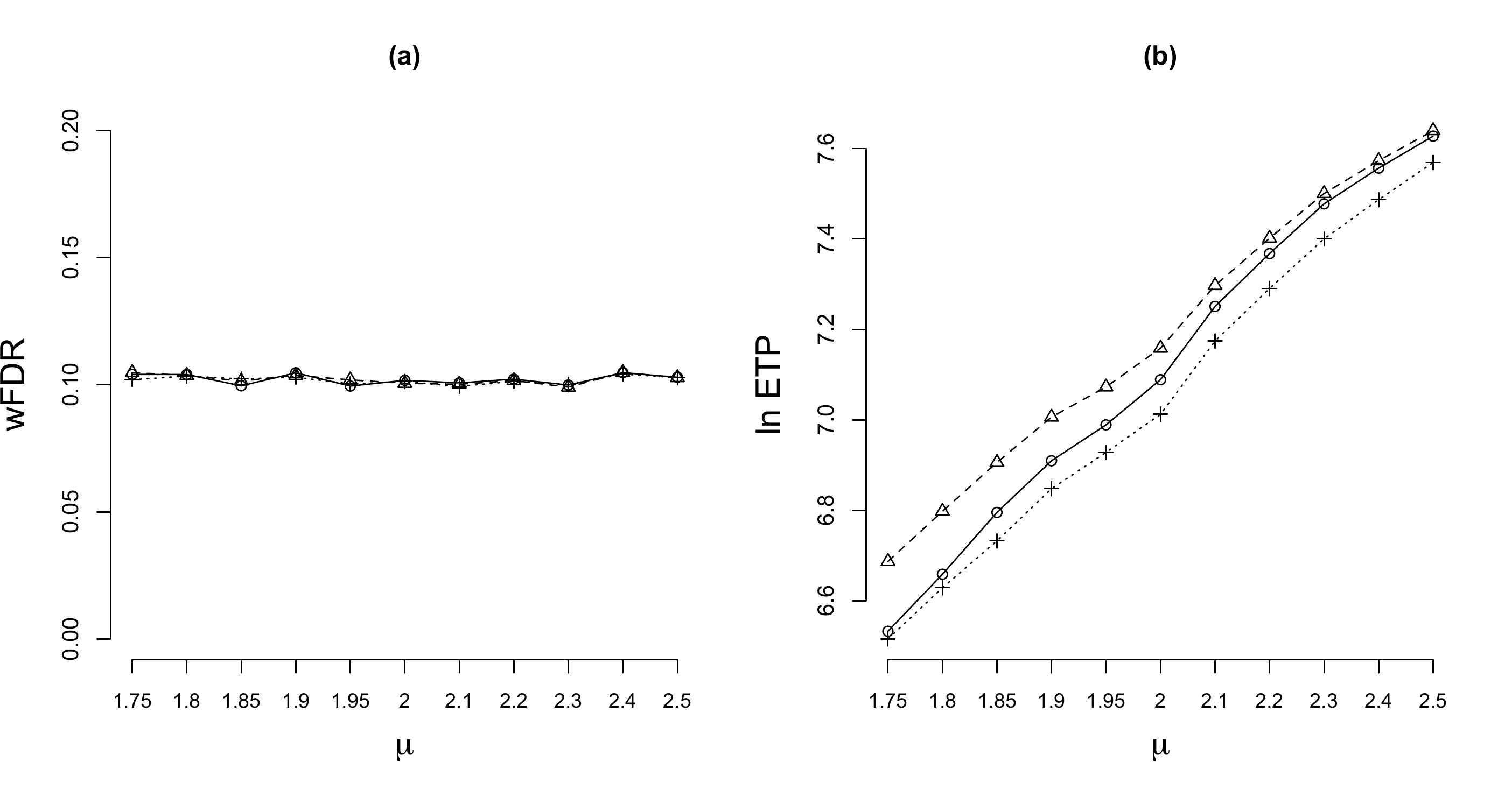} 
\includegraphics[width=\textwidth, height=2.6in]{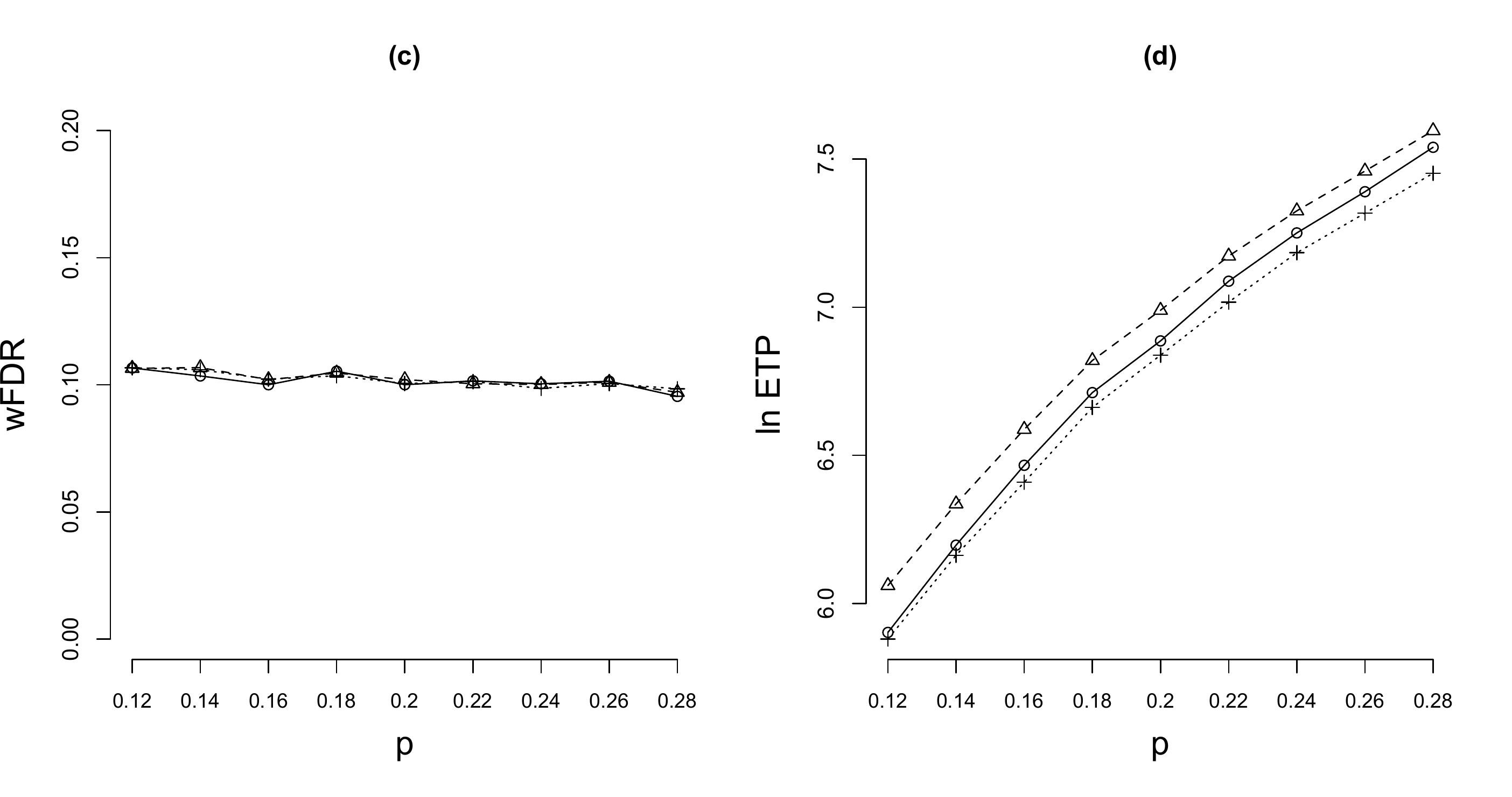} 
\includegraphics[width=\textwidth, height=2.6in]{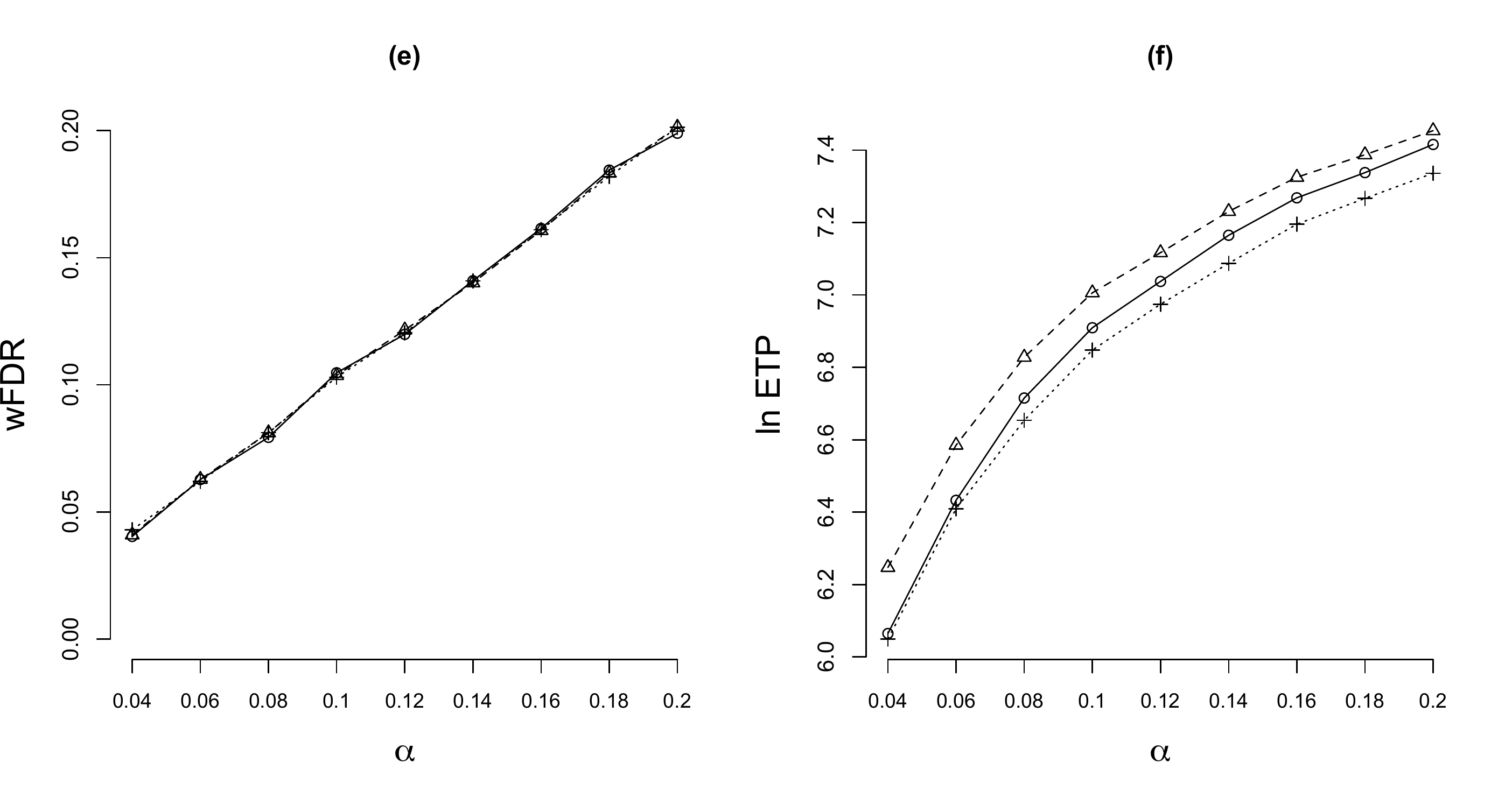} 
\caption{Comparison with general weights: $\circ$ WPO, $\triangle$ DD (proposed), and $+$ AZ. All methods control the wFDR approximately at the nominal level. The efficiency gains of the proposed method become more pronounced when (i) the signal strength decreases, (ii) the signals become more sparse, or (iii) the test level $\alpha$ decreases.}
\label{figure3_plot}
\end{figure}

In the last simulation study, $a_i$'s are assigned to two groups of hypotheses with group sizes $m_1 = 3000$ and $m_2 = 1500$. In groups 1 and 2, we fix $a_i=1$ and $a_i=3$, respectively. Conventional FDR methods are only guaranteed to work when all $a_i$ are fixed at 1. Under this setting, we expect that the unweighted AZ may fail to control the wFDR. We then generate random weights $b_i$ from log-normal distribution with location $\ln 6$ and scale $1$. The non-null proportion for group 1 is 0.2, and that for group 2 is 0.1. The mean of the the non-null distribution for group 1 or $\mu_1$ is varied between $[-3.75, -2]$ while that for group 2 is fixed at 2. The simulation results are shown in Figure \ref{figure4_plot}. We can see that the unweighted AZ method fails to control the wFDR at the nominal level, which verifies our conjecture. The observations regarding the ETP are similar to those in the previous simulation study. Overall, all numerical studies together substantiate our theoretical results and affirm the use of the methodology in various settings. 

\begin{figure}
\centering
\includegraphics[max size={\textwidth}{\textheight}]{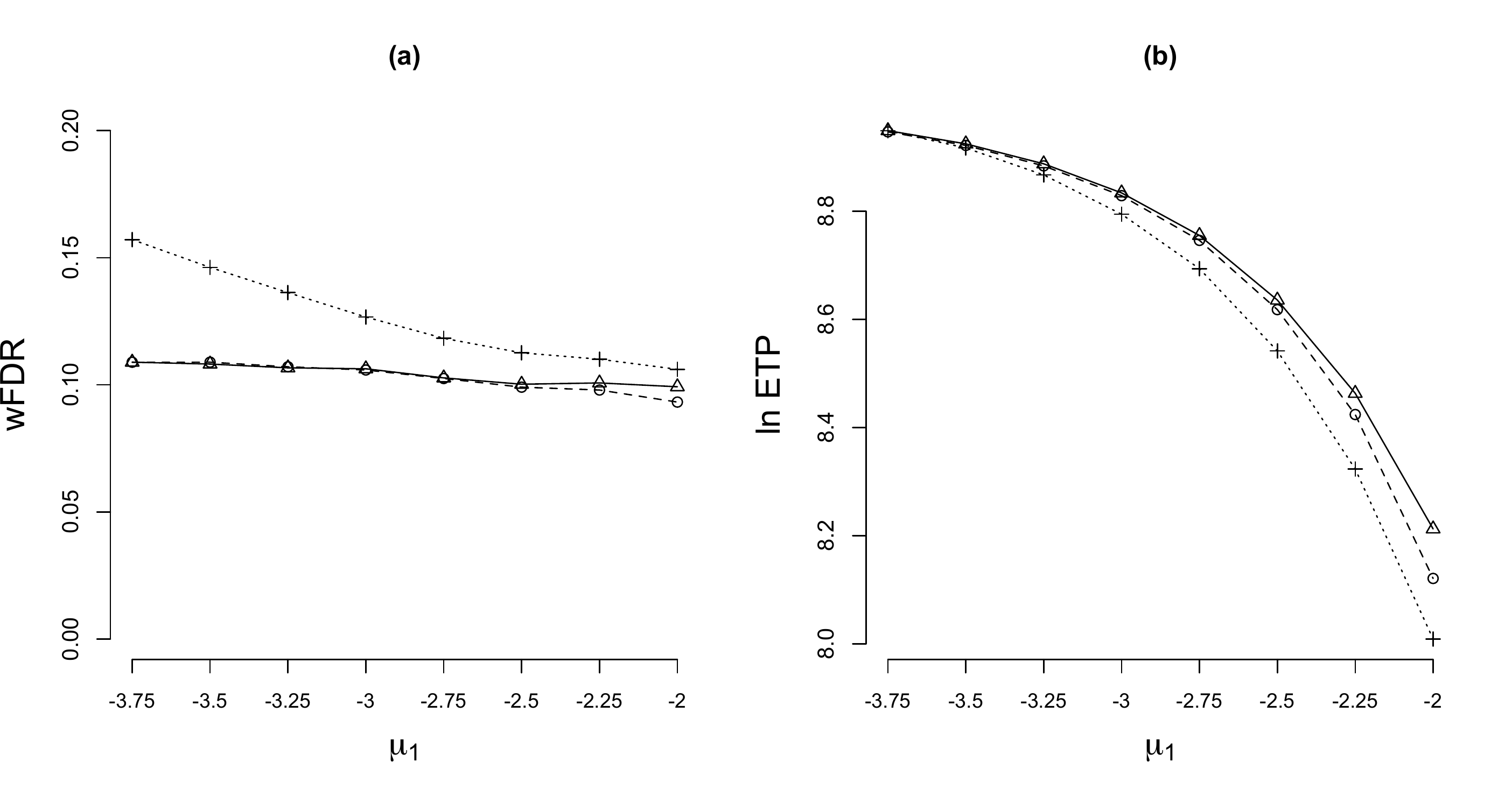} 
\caption{Comparison with general weights: $\circ$ WPO, $\triangle$ DD (proposed), and $+$ AZ. The unweighted AZ method fails to control the wFDR at the nominal level. The efficiency gain of the proposed method increases as signals become weaker.}
\label{figure4_plot}
\end{figure}

\subsection{Comparison of the wFDR definitions}

Section 2.2 introduces two wFDR definitions (2.4) and (2.5). It is shown in Propositions \ref{lemma_weight_eqv} and \ref{weight_eqv_dd} that (2.4) and (2.5) are asymptotically equivalent. Next we conduct a small simulation study to compare the two definitions in finite samples.

We generate $z$-values from model (5.1) with $\mu=1.9$, $\sigma=1$ and $p=0.2$.  The weights $a_i$ are generated from a log-normal distribution with $\mu=\log 3$ and $\sigma=1$, $b_i$ are chosen as 1 for all $i$. To see how quickly the asymptotics kick in, we vary the number of tests from 500 to 5000 and apply two wFDR procedures, namely the BH97 and DD, to the simulated data. Figure \ref{wFDR-defs.fig} summarizes the comparison results of two wFDR definitions, which are computed by averaging the results from 200 replications. Panel (a) considers the case when the BH97 method is implemented. We can see that (2.5), the proposed wFDR definition, is slightly higher than (2.4), the wFDR definition in Benjamini and Hochberg (1997). However, the two wFDR levels quickly become very similar when $m$ increases to about 3000, an order of magnitude that is easily fulfilled by large-scale studies such as GWAS. Panel (b) considers the case when Procedure 1 is implemented. We can see that the two definitions already agree  very well  when $m$ is only 500. These numerical results support our claim based on theoretical analysis that the two definitions are asymptotically equivalent. 

\begin{figure}[h]
\includegraphics[width=6.2in, height=4in]{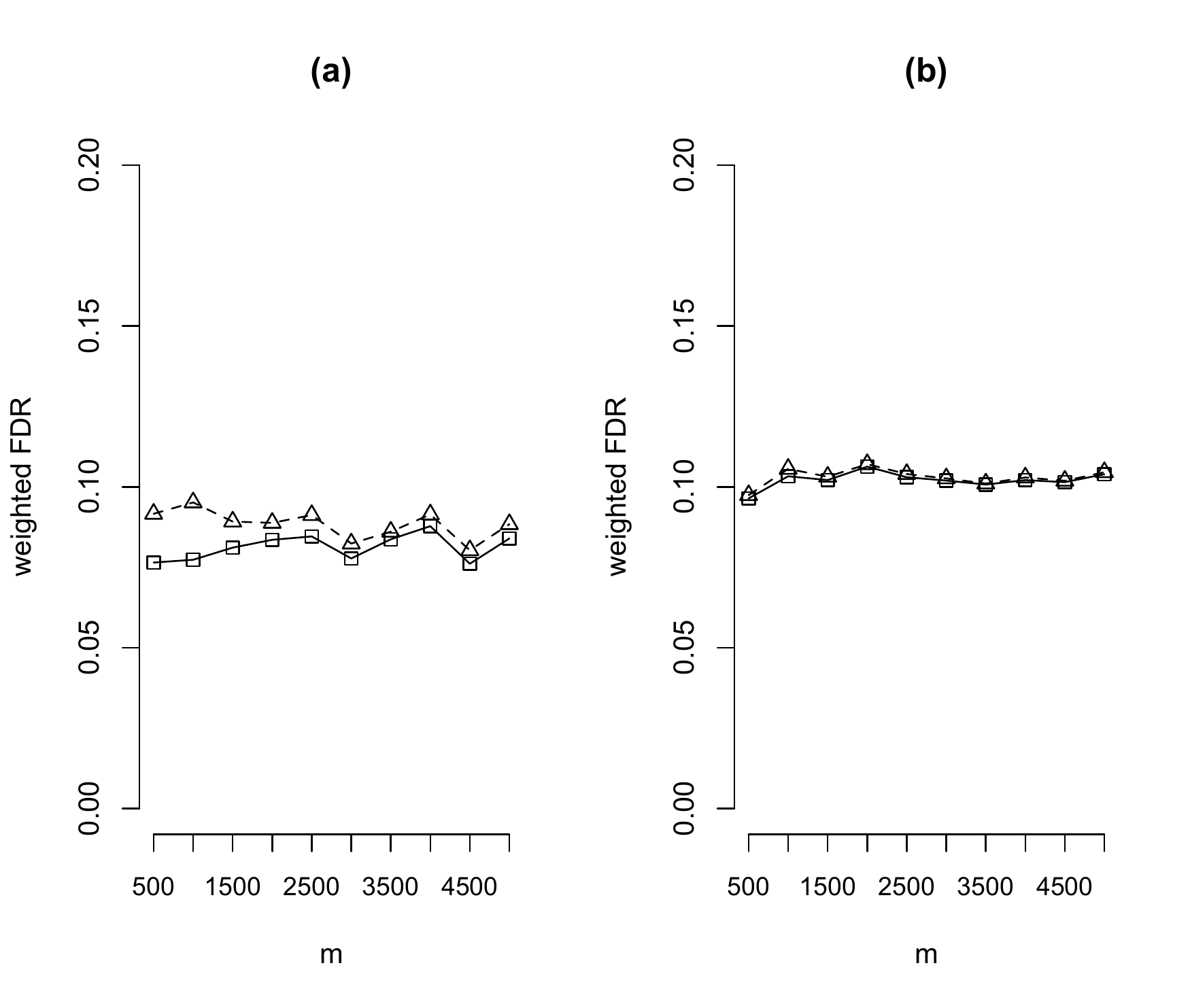}
\caption{Illustration of the finite sample approximation of (a) BH97, and (b) the proposed data driven procedure where $\Box$ indicates $wFDR_{BH}$ definition and $\triangle$ indicates proposed $wFDR$ definition.}
\label{wFDR-defs.fig}
\end{figure}

\subsection{Finite sample performance of the proposed estimator}

In the literature, different methods have been proposed to estimate the Lfdr statistics. We compare two approaches: the method using R-package ``locfdr'' and the method in Sun and Cai (2007). The key difference of the two methods is in the estimation of the non-null proportion. The issue is briefly mentioned in Remark 2; see Jin and Cai (2007) for details. We use the same model and same weights as the previous section. The MSEs of the Lfdr and ranking statistics for varying $m$ and $p$ are reported in Table \ref{lfdr-comp.tab}. We can see that the method in Sun and Cai (2007) has smaller MSEs in all settings considered. This method is used in all our simulation studies. For a more detailed comparison in the multiple testing setting, see Cai and Jin (2010). 

\begin{table}\caption{Comparisons of RMSE of various methods. Results are reported as 100*mean(100*sd) of 100 replications.}\label{lfdr-comp.tab}
\begin{center}
\begin{tabular}{c|cccccc}
\multicolumn{7}{c}{}\\
\hline
p=? & 0.1 & 0.1 & 0.15 & 0.15 & 0.2 & 0.2\\
\hline
\multicolumn{7}{c}{Local fdr statistic}\\
\hline
m? & Our & locfdr & Our & locfdr & Our & locfdr \\
\hline
1000 & 6.25(2.19) & 12.8(2.40) & 5.71(1.72) & 19.5(2.31) & 6.01(1.81) & 26.7(2.66)\\
2000 & 4.90(1.47) & 11.8(1.82) & 4.67(1.46) & 17.7(2.03) & 4.64(1.35) & 25.2(2.12)\\
5000 & 3.75(1.29) & 10.7(1.58) & 3.47(1.02) & 16.3(1.43) & 3.63(1.03) & 23.4(1.58)\\
\hline
\multicolumn{7}{c}{Ranking Statistic}\\
\hline
1000 & 5.54(1.78) & 11.9(2.26) & 4.86(1.45) & 19.1(2.41) & 5.04(1.44) & 27.3(2.46)\\
2000 & 4.29(1.21) & 11.2(1.80) & 3.94(1.07) & 17.6(2.22) & 3.96(1.06) & 26.1(2.01)\\
5000 & 3.18(1.03) & 10.2(1.65) & 3.03(0.81) & 16.6(1.52) & 3.10(0.79) & 24.8(1.68)\\
\hline
\end{tabular}
\end{center}
\end{table}

\subsection{Impacts of weights}

Our previous simulation studies show that the proposed method is optimal when the weighted power function is used to assess the performance. This section investigates the performance of the proposed method when the \emph{unweighted power function} is used to assess the performance. It is important to note that our ultimate goal is not to improve the power (although power may be gained as a byproduct when weights are informative, as shown by the simulation below) but to make more informed decisions. Hence this comparison is only of conceptual interest. 

In our simulation, $m=3000$ $z$-values are generated from model (5.1). Let $\pmb s=(s_1, \cdots, s_m)$ denote external covariates that follow a log normal distribution with a mean of -1.5 and variance of 1. The probability that a case is a non-null is given by $p_i=\mbox{P}(\theta_i=1)=s_i/(s_i+1)$. We set $a_i=1$ for all $i$. The power functions are modified to incorporate the external information. To investigate the impact of the weights, we consider three settings: 
\begin{description}
\item (i) highly informative weights: $b_i=(1+s_i)^{1/2}$.
\item (ii) moderately informative weights: $b_i=(1+s_i)^{1/8}$.
\item (iii) anti-informative weights: $b_i=(1+s_i)^{-1/8}$.
\end{description}
To compare the ranking efficiencies (with the unweighted power) of different methods, we compute the ETP in top $k$ hypotheses ($k=100, 200, 300$). The following methods are considered in our comparison:
\begin{itemize}
\item BH97 (denoted $\square$, Benjamini and Hochberg 1997).
\item Lfdr ($+$, Sun and Cai 2007).
\item DD ($\triangle$, proposed). 
\end{itemize}
We simulate 200 data sets and plot the averages of the log of the ETP as functions of $\mu$; the results are summarized in Figure \ref{impact-weights.fig}. We can see that the proposed method has the largest power when the weights are informative, and the efficiency gain is large when the signals are weak. When the weights are anti-informative, the proposed method is less powerful than the Lfdr method but still dominates the BH97 method in most settings. The findings are similar to the tradeoffs reported in Genovese et al.~(2006). 

\begin{figure}[h]
\includegraphics[width=6in]{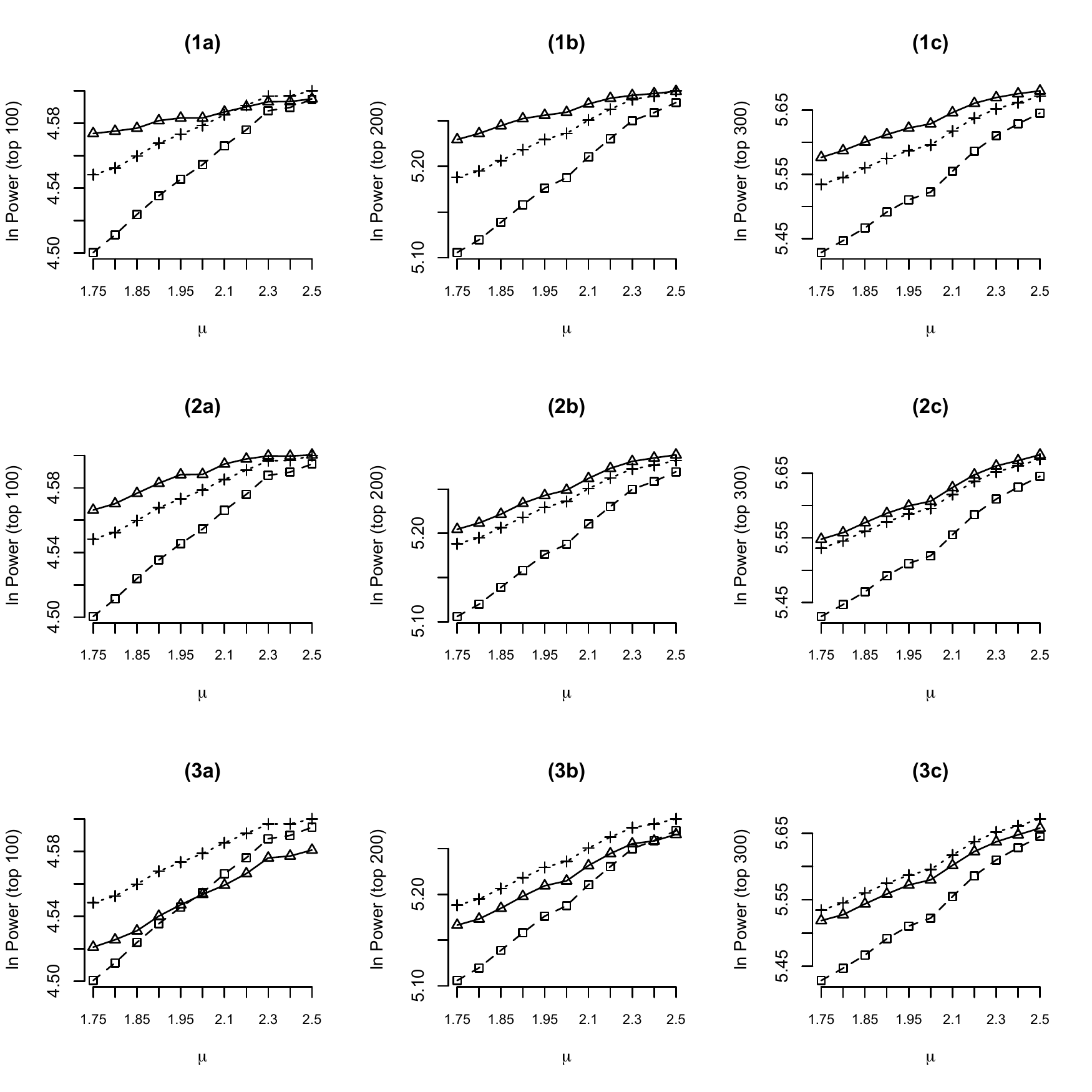}
\caption{The comparison of BH97 $\square$, Lfdr $+$, and DD $\triangle$. The top, middle and bottom rows correspond to the cases with highly informative, moderate informative and anti-informative weights.}
\label{impact-weights.fig}
\end{figure}

\end{document}